\documentclass[10pt,aps,prd,twocolumn,floats,floatfix]{revtex4-2}
\usepackage{graphicx}
\usepackage{dcolumn}
\usepackage{bm}
\usepackage{amsmath}
\usepackage{amssymb}
\usepackage{amsthm}
\usepackage{mathrsfs}
\usepackage{xcolor}
\usepackage{multirow}
\usepackage{comment}
\usepackage[colorlinks=true,linkcolor=blue,citecolor=blue,urlcolor=blue]{hyperref}

\begin{document}

\title{Covariant phase space analysis of static spherical vacua in power-law \texorpdfstring{$f(R)$}{f(R)} gravity}

\author{Mariam Campbell${}^{1}$} \email{CampbellM@ukzn.ac.za}
\author{Eduardo Bittencourt${}^{2,3}$}\email{bittencourt@unifei.edu.br}
\author{Peter K. S. Dunsby${}^{3,4,5}$}\email{peter.dunsby@uct.ac.za}
\affiliation{${}^{1}$Astrophysics Research Centre, School of Agriculture and Science, University of KwaZulu-Natal, Private Bag X54001, Durban 4000, South Africa}
\affiliation{${}^{2}$Institute of Physics and Chemistry, Federal University of Itajub\'a, Av.\ BPS 1303, Itajub\'a/MG, Brazil}
\affiliation{${}^{3}$Department of Mathematics and Applied Mathematics, University of Cape Town, Rondebosch 7700, Cape Town, South Africa}
\affiliation{${}^{4}$Center for Space Research, North-West University, Potchefstroom 2520, South Africa}
\affiliation{${}^{5}$South African Astronomical Observatory, Observatory 7925, Cape Town, South Africa}
\date{\today}

\begin{abstract}
We formulate the static, spherically symmetric vacuum equations of power-law metric $f(R)$ gravity as a covariant autonomous system using the $(1+1+2)$ formalism. The finite critical sets include the scale-invariant Clifton--Barrow solution, a Minkowski endpoint shared by the Schwarzschild family, and degenerate boundaries of the normalized chart. The Clifton--Barrow equilibrium is an outward-radial attractor over a positive interval of the power-law parameter, whereas the Schwarzschild endpoint is a saddle of the polynomially extended system. Poincar\'e compactification, supplemented by secondary and weighted charts, resolves several asymptotic regimes. In particular, normally attracting sheets at large normalized curvature occur in the positive-curvature sector and are separated from the negative-curvature Clifton--Barrow equilibrium. Metric reconstruction also identifies a selected scalar-flat family with a regular throat, one asymptotically flat end, and a curvature-singular end. These results organize the regular reduced chart and the boundary regimes resolved here; they do not constitute a complete classification of static spherical vacua. All stability statements concern the static radial flow.
\end{abstract}

\maketitle

\section{Introduction}
Metric $f(R)$ gravity replaces the Ricci scalar in the Einstein--Hilbert action by a nonlinear function and provides a simple setting in which to study additional gravitational degrees of freedom \cite{Buchdahl1970,SotiriouFaraoni2010,DeFeliceTsujikawa2010,Clifton:2011jh}. Its applications include cosmological acceleration and inflation \cite{Turner:2022gvw}. For generic nonlinear $f(R)$, the scalar degree of freedom makes the metric equations fourth order and enlarges the vacuum solution space, while also making exact integration more difficult.

Static spherical vacua provide a useful setting for examining this additional freedom. Unlike in general relativity (GR), the vacuum scalar curvature need not be constant, and the usual Birkhoff conclusion requires additional assumptions \cite{Faraoni:2010rt,Nzioki:2009av,Nzioki:2013lca}. Direct reductions, reconstruction methods, and geometric restrictions have produced several classes of spherical solutions \cite{MultamakiVilja2006,CapozzielloStabileTroisi2008,SebastianiZerbini2011,GuilabertEtAl2024}. For the power-law model, the Clifton--Barrow family supplies an exact benchmark \cite{CliftonBarrow2005,Clifton:2006ug}. Earlier dynamical analyses already identified its scale-invariant large-radius attractor and radial perturbation modes \cite{MignemiWiltshire1989,CliftonBarrow2005}. The full family has a Schwarzschild limit as the power-law parameter tends to zero. This limit must be distinguished from imposing $R=0$ at fixed positive parameter, where $f'(0)=0$ and the Einstein-like curvature-fluid representation is undefined.

The $(1+1+2)$ formalism expresses spherical geometry in covariantly defined scalars \cite{ClarksonBarrett2003,BetschartClarkson2004,Clarkson2007}. Applications include Schwarzschild perturbations \cite{ClarksonBarrett2003}, electromagnetic black-hole ringdown \cite{Clarkson:2003mp}, stellar structure with single and multiple fluids \cite{Carloni-generating-theorems1,Carloni-generating-theorems2,Naidu:2021yfz}, and stellar models in modified gravity \cite{Campbell:2024fRstars,Bittencourt2026a,Bittencourt2026b}. Here we use the static vacuum equations developed by Nzioki et al. \cite{Nzioki:2009av} to obtain a three-dimensional normalized system for the signed power-law model.

Our contribution is a Jordan-frame covariant description that connects the finite critical sets, the resolved directions at infinity, and their metric reconstructions. It reproduces the known Clifton--Barrow radial spectrum and makes explicit the separation of the two curvature-sign sectors. Secondary charts retain the full vector field where the leading compactification degenerates. We also reconstruct selected scalar-flat boundary orbits and identify their relation to the Buchdahl-inspired metrics \cite{Nguyen2023Buchdahl,NguyenAzregAinou2023Wormholes}. The Schwarzschild family is treated as a separately selected Ricci-flat extension, since the exact scalar-flat sector is degenerate.

Throughout, stability means stability of the static radial flow as the areal radius increases. It does not mean stability against time-dependent perturbations. Most geometric conclusions concern positive values of the power-law parameter; the generic reduced chart excludes the GR limit and other exceptional cases, which must be considered separately.

Section~\ref{sec:Prop_eqs} reviews the propagation equations, and Sec.~\ref{sec:vac_prop_eqs} introduces the normalized vacuum system. Section~\ref{subsec:qualitative_reduced_system} analyzes the power-law system and its compactification. Section~\ref{sec:metric_reconstruction_fixed_points} reconstructs the associated metrics and relates them to known solutions. Appendix~\ref{app:purple_scalarflat} supplies the scalar-flat curvature checks, the regular throat extension, and the comparison with the Buchdahl-inspired family. Section~\ref{sec:conclusion} summarizes the results.

\section{Propagation equations}\label{sec:Prop_eqs}
We work in the Jordan-frame metric formulation of \(f(R)\) gravity \cite{SotiriouFaraoni2010,DeFeliceTsujikawa2010}, with action
\begin{equation}
S=\frac{1}{2}\int d^4x\sqrt{-g}\,f(R)+S_{\rm m}[g_{ab},\Psi_{\rm m}] ,
\end{equation}
where \(8\pi G=c=1\), \(R\) is the four-dimensional Ricci scalar, and \(\Psi_{\rm m}\) collectively denote the matter fields. The connection is taken to be the Levi--Civita connection of \(g_{ab}\) and is not varied independently. Variation with respect to \(g^{ab}\) therefore gives
\begin{equation}
\label{eq:fR_field_eqs}
f' R_{ab}-\frac{1}{2}f g_{ab} -\nabla_a\nabla_b f' +g_{ab}\Box f' = T^{\rm m}_{ab},
\end{equation}
where prime means the derivative with respect to $R$. Equivalently, for $f'\neq0$, these equations may be written in the Einstein-like form as follows:
\begin{equation}\label{eq:Einstein_like_fR}
G_{ab}=T^{\rm tot}_{ab}, \qquad T^{\rm tot}_{ab} = \frac{1}{f'}T^{\rm m}_{ab} + T^{R}_{ab},
\end{equation}
where the effective curvature-fluid energy-momentum tensor is
\begin{equation}
\label{eq:curv_T_ab}
T^{R}_{ab} = \frac{1}{f'} \left[ \frac{1}{2}\left(f-Rf'\right)g_{ab} +\nabla_a\nabla_bf' -g_{ab}\Box f' \right] .
\end{equation}
Thus, the field equations of $f(R)$ gravity can be interpreted as GR sourced by two effective components: the rescaled matter sector $T^{\rm m}_{ab}/f'$ and the curvature sector $T^R_{ab}$.

We now specialize to static locally rotationally symmetric class~II (LRS-II) spacetimes and employ the \(1+1+2\) decomposition \cite{ClarksonBarrett2003,BetschartClarkson2004,Clarkson2007,Nzioki:2009av}. Let the unit timelike vector be denoted by \(u^a\), while \(e^a\) defines the preferred radial direction, with 
\begin{equation}
u^a u_a=-1,\qquad e^a e_a=1,\qquad u^a e_a=0.
\end{equation}
The metric induced on the local rest spaces of $u^a$ is
\begin{equation}
h_{ab}=g_{ab}+u_a u_b,
\end{equation}
while the metric induced on the two-sheets orthogonal to both $u^a$ and $e^a$ is
\begin{equation}
N_{ab}=h_{ab}-e_a e_b = g_{ab}+u_a u_b-e_a e_b .
\end{equation}
For any scalar $\psi$, the dot and hat derivatives are defined by
\begin{equation}
\dot\psi=u^a\nabla_a\psi, \qquad \hat\psi=e^aD_a\psi, \qquad D_a\psi=h_a{}^b\nabla_b\psi .
\end{equation}
In a static LRS-II spacetime, all scalars have vanishing dot derivatives and no sheet gradients. The nontrivial geometrical scalars are the sheet expansion $\phi$, the acceleration scalar $A$, and the electric Weyl scalar $\mathcal E$, defined by
\begin{equation}
\phi=N_a{}^{b}D_b e^a, \quad A=e^a\dot u_a, \quad E_{ab}=\mathcal E\left(e_a e_b-\frac12N_{ab}\right),
\end{equation}
where $E_{ab}=C_{acbd}u^cu^d$ is the electric part of the Weyl tensor.

The effective total energy-momentum tensor is decomposed as
\begin{eqnarray}
T^{\rm tot}_{ab} &=& \mu^{\rm tot}u_a u_b +p^{\rm tot}h_{ab} +2Q^{\rm tot}u_{(a}e_{b)}\nonumber\\[1ex]
&&+\Pi^{\rm tot}\left(e_a e_b-\frac12N_{ab}\right).\label{eq:Ttot_decomp}
\end{eqnarray}
In a static LRS-II region, with \(u^a\) aligned with the timelike Killing congruence, the mixed \(u^a e^b\) projection of the field equations vanishes, and consequently \(Q^{\rm tot}=0\). Thus, the absence of an effective radial energy flux follows from staticity in the Killing frame. The remaining total effective variables are
\begin{align}
\mu^{\rm tot} &= \frac{\mu^{\rm m}}{f'}+\mu^R,\\[1ex]
p^{\rm tot} &= \frac{p^{\rm m}}{f'}+p^R,\\[1ex]
\Pi^{\rm tot} &= \frac{\Pi^{\rm m}}{f'}+\Pi^R .
\end{align}
Even when the material source is a perfect fluid, \(\Pi^{\rm m}=0\), the curvature sector generally produces a nonzero effective anisotropic stress \(\Pi^R\).

Since the spacetime is static, the Ricci scalar depends only on the radial direction defined by $e^a$. So, it is useful to introduce $X\equiv \hat R$, such that 
\begin{equation}
\widehat{f'}=f''X, \qquad \widehat{\widehat{f'}}=f''\widehat X+f'''X^2 .
\end{equation}
Then, the curvature-fluid variables obtained by projecting Eq.~\eqref{eq:curv_T_ab} with respect to $u^a$, $e^a$, and $N_{ab}$ can be written as \cite{Nzioki:2009av}
\begin{align}
\mu^R &= \frac{1}{f'}\left[ \frac12(Rf'-f)+f''(\hat X+\phi X)+f'''X^2\right], \label{eq:muR}\\
p^R &= \frac{1}{f'}\left[ \frac12(f-Rf')-\frac23(f''\hat X+f'''X^2) \right.\nonumber\\[-0.5ex]
    &\hspace{3em}\left.-\left(\frac23\phi+A\right)f''X\right], \label{eq:pR}\\
\Pi^R &= \frac{2}{3f'}\left[ f''\hat X+f'''X^2-\frac12\phi f''X\right]. \label{eq:PiR}
\end{align}

These equations show explicitly why the curvature sector behaves as an effective anisotropic fluid, with effective energy density and isotropic pressure that are not necessarily positive definite. Nevertheless, the curvature fluid is a geometrical bookkeeping device. Negative effective density or violations of effective-fluid energy conditions do not by themselves identify ghosts, since the fundamental conditions involve $f'$, $f''$, and perturbations of the scalar degree of freedom.

The trace of Eq.~\eqref{eq:fR_field_eqs} gives the scalaron equation
\begin{equation}
\label{eq:fR_trace_general}
f'R-2f+3\Box f'=-\mu^{\rm m}+3p^{\rm m}.
\end{equation}
In the case of static LRS-II spacetimes, one obtains
\begin{equation}
\Box f'=\widehat{\widehat{f'}}+(\phi+A)\widehat f',
\end{equation}
and therefore
\begin{equation}\label{eq:fR_trace_LRS}
f'R-2f +3\left[ f''\hat X+f'''X^2+(\phi+A)f''X \right] = -\mu^{\rm m}+3p^{\rm m}.
\end{equation}
When $f''\neq0$, Eq.~\eqref{eq:fR_trace_LRS} may be used as a propagation equation for $X$:
\begin{equation}\label{eq:X_prop}
\hat X = \frac23\frac{f}{f^{\prime\prime}}-\frac13 \frac{f^{\prime}}{f^{\prime\prime}}R - \frac{f^{\prime\prime\prime}}{f^{\prime\prime}}X^{2}- X( \phi +A )-\frac{\mu^{\rm m}-3p^{\rm m}}{3f''}.
\end{equation}
If \(f''\equiv0\), then \(f(R)=aR+b\). For \(a\neq0\), this is GR with a rescaled gravitational coupling, \(G_{\rm eff}=G/a\), and cosmological constant \(\Lambda=-b/(2a)\). The case \(a=0\) is degenerate and contains no Einstein–Hilbert term. If \(f''\) vanishes only at isolated values of \(R\), Eq.~\eqref{eq:X_prop} is not defined there, and the original undivided trace equation must be retained.

The independent $(1+1+2)$ propagation equations for the static LRS-II geometry are then
\begin{align}
&\hat\phi = -\frac12\phi^2 -\frac23\mu^{\rm tot} -\frac12\Pi^{\rm tot} -\mathcal E,\label{eq:hat_phi}\\[1ex]
&\hat{A} =-\left(A+\phi\right)A +\frac12\left(\mu^{\rm tot}+3p^{\rm tot}\right),\label{eq:hat_A}\\[1ex]
&\hat p^{\rm tot} +\hat\Pi^{\rm tot} = -\left(\frac32\phi+A\right)\Pi^{\rm tot} -\left(\mu^{\rm tot}+p^{\rm tot}\right)A,\label{eq:tot_conservation}\\[1ex]
&\hat{\mathcal E} -\frac{1}{3}\hat\mu^{\rm tot} +\frac12\hat\Pi^{\rm tot} = -\frac32\phi \left( \mathcal E+\frac12\Pi^{\rm tot}\right),\label{eq:hat_E}\\[1ex]
&0 = -A\phi +\frac13\left(\mu^{\rm tot}+3p^{\rm tot}\right) -\mathcal E +\frac12\Pi^{\rm tot}.\label{eq:mixed_constraint}
\end{align}
Equation~\eqref{eq:tot_conservation} is the radial conservation equation for the total effective source, whereas Eq.~\eqref{eq:mixed_constraint} is an algebraic constraint relating the acceleration, the Weyl curvature, and the effective fluid variables. Eliminating \(\mathcal E\) through Eq.~\eqref{eq:mixed_constraint} and using the remaining propagation equations makes Eq.~\eqref{eq:hat_E} redundant. It therefore need not be evolved independently. 

For metric reconstruction, it is useful to introduce the Gaussian curvature \(K\) of the two-sphere symmetry, which satisfies
\begin{equation}\label{eq:K_constraint}
K = \frac13\mu^{\rm tot} -\mathcal E -\frac12\Pi^{\rm tot} +\frac14\phi^2.
\end{equation}

For completeness, a minimally coupled material sector satisfies \(\nabla^bT^{\rm m}_{ab}=0\). Since the remainder of this work is restricted to vacuum, no material equation of state or additional closure relation is required; all material variables shall be set to zero in the forthcoming sections.

The definition of \(X\), together with the trace equation~\eqref{eq:X_prop}, supplies the radial propagation of the curvature variables. The remaining propagation equations are \eqref{eq:hat_phi}--\eqref{eq:tot_conservation}, supplemented by the constraints \eqref{eq:mixed_constraint} and \eqref{eq:K_constraint}. The curvature-fluid variables entering the total effective source are given by \eqref{eq:muR}--\eqref{eq:PiR}.

Because Eq.~\eqref{eq:X_prop} was obtained by division by \(f''\), the general-relativistic limit cannot be taken directly in that propagation equation. It must instead be taken in the original undivided field equations, Eq.~\eqref{eq:fR_field_eqs}, or equivalently in the trace equation~\eqref{eq:fR_trace_general}. Setting
\begin{equation}
f(R)=R-2\Lambda, \qquad f'=1, \qquad f''=0,
\end{equation}
all derivative terms in the curvature fluid vanish and one obtains
\begin{equation}
\mu^R=\Lambda, \qquad p^R=-\Lambda, \qquad \Pi^R=0.
\end{equation}
For $\Lambda=0$, the standard static LRS-II propagation equations of GR are recovered.

\section{General normalized vacuum system}\label{sec:vac_prop_eqs}
Following the $(1+1+2)$ covariant splitting introduced by Nzioki et al. \cite{Nzioki:2009av}, the field equations for static, spherically symmetric spacetimes in metric $f(R)$ gravity reduce to a system of spatial propagation equations along the radial vector field $e^a$. There, this structure is employed to examine metric reconstruction algorithms and the non-uniqueness of static vacuum states. Here we convert the propagation equations into an autonomous system and compactify its generic reduced chart in order to organize both its finite critical sets and its asymptotic boundaries. Setting the matter energy-momentum tensor to zero ($T^{\rm m}_{ab} = 0$), Eq.~\eqref{eq:tot_conservation} becomes a differential consequence of the metric field equations through the contracted Bianchi identity; it is not an independent propagation equation for the curvature-fluid representation.

To analyze the dynamics, we introduce the following normalized variables 
\begin{align}
&\bar R\equiv \frac{R}{\phi^2}, \qquad \bar X\equiv \frac{X}{R\, \phi}, \qquad \bar A\equiv \frac{A}{\phi}, \nonumber\\
&\bar p\equiv\frac{p^{R}}{\phi^{2}}, \qquad \bar\Pi\equiv\frac{\Pi^{R}}{\phi^{2}}, \qquad \bar\mu\equiv\frac{\mu^{R}}{\phi^{2}},\label{norm_vars}\\
&\Xi\equiv \frac{\phi_{,\rho}}{\phi}, \qquad \bar K\equiv \frac{K}{\phi^2}, \qquad \bar Z\equiv\frac{X}{\phi^3}=\bar R\bar X,\nonumber
\end{align}
For any scalar \(\psi\), the radial parameter \(\rho\) is defined through the parametrization
\begin{equation}
    \label{eq:hat_rho_der_rel} \widehat{\psi}=\phi\,\psi_{,\rho}.
\end{equation}
Its relation to the area radius \(r\) is given by \cite{Campbell:2024fRstars}
\begin{equation}
\label{eq:rho_r_rel}  
\rho=2\ln\left(\frac{r}{r_0}\right).
\end{equation}
Consequently, \(\rho\to+\infty\) corresponds to increasing area radius, and a linear mode \(e^{\lambda\rho}\) has the radial scaling
\[
e^{\lambda\rho} =\left(\frac{r}{r_0}\right)^{2\lambda}.
\]
Stability with respect to increasing \(\rho\) therefore means outward-radial stability and must not be confused with stability against time-dependent perturbations.

The variables also have useful interpretations:
\[
\bar X  =\frac{d\ln|R|}{d\rho} =\frac12\frac{d\ln|R|}{d\ln r}, \qquad \Xi=\frac{d\ln|\phi|}{d\rho}.
\]
Within the present chart, \(\bar Z=\bar R\,\bar X\) is a derived rather than an independent variable. Its direct definition, \(\bar Z=X/\phi^3\), is nevertheless useful when studying limits toward \(R=0\), where \(\bar X\) itself is not defined.

We assume that \(f\) is at least \(\mathcal C^3\) on the curvature interval under consideration. The normalized chart requires
\[
\phi\neq0,\qquad R\neq0,\qquad f'(R)\neq0,\qquad f''(R)\neq0.
\]
Boundaries at which any of these conditions fail must be treated using different variables or directly from the original undivided
field equations.

Since an $f(R)$ theory can be rearranged into the Einstein-like form \eqref{eq:Einstein_like_fR} on regions where \(f'(R)\neq0\), this allows us to directly translate Eqs. \eqref{eq:X_prop}-\eqref{eq:mixed_constraint} as a single-fluid system, in which the curvature fluid encapsulates the geometrical source terms. The curvature fluid is directly related to the kinematical quantities of the system, and this significantly reduces the set of independent equations to
\begin{align}
    \bar{A}_{,\rho}=&-\bar{A}\left(1+\Xi+\bar{A}\right)+\frac{1}{2}\left(\bar{\mu}+3\bar{p}\right),\label{eq: dAbar general f}\\
    \bar{R}_{,\rho}=&\left(\bar{X}-2\Xi\right)\bar{R},\label{eq: dRbar general f}\\
    \bar{X}_{,\rho}=& -\left(1+\bar{A}+\bar{X}+\Xi\right)\bar{X}+\frac{2f(R) - \phi^{2}\bar{R}f'(R)}{3\phi^{4}\bar{R}f''(R)}\nonumber\\
     &-\frac{\phi^{2}\bar{R}\bar{X}^{2}f'''(R)}{f''(R)}.\label{eq: dXbar general f}
\end{align}

For a generic specified function \(f(R)\), the last two terms of Eq.~\eqref{eq: dXbar general f} must be evaluated at $R=\phi^2\bar R$. They therefore retain information about the absolute curvature scale and cannot, in general, be expressed solely in terms of \((\bar A,\bar R,\bar X)\). The variable \(\phi\) must consequently be retained. Using Eq.~\eqref{eq:hat_phi} together with the constraint~\eqref{eq:mixed_constraint} gives
\begin{equation}
    \frac{\phi_{,\rho}}{\phi} \equiv\Xi = \bar{A}-\bar{\mu}-(\bar{p}+\bar{\Pi})-\frac{1}{2}.
    \label{eq: Xi general}
\end{equation}
Thus, Eqs.~\eqref{eq: dAbar general f}--\eqref{eq: Xi general} provide the normalized propagation structure. Closure requires expressing \(\bar\mu,\bar p,\bar\Pi\) in terms of the state variables for a specified function \(f(R)\). The Gaussian curvature is not an additional dynamical variable. In fact, using Eq.~\eqref{eq:K_constraint}, it is reconstructed algebraically as
\begin{equation}
\bar{K} = \frac{1}{4}+\bar{A}-(\bar{p}+\bar{\Pi}).\label{eq: Kbar general}
\end{equation}
Nevertheless, its radial evolution 
\begin{equation}
\bar K_{,\rho}=-(1+2\Xi)\bar K,
\end{equation}
together with Eq.~\eqref{eq: dRbar general f}, separates the dynamics into disjoint sign sectors since the signs of $\bar K$ and $\bar R$ are preserved wherever the normalized chart is regular.

For genuine spherical two-sheets of finite area, \(K>0\). Since \(\phi^2>0\), the physical spherical sector therefore satisfies \(\bar K>0\). The set \(\bar K=0\) is a boundary or asymptotic limit and does not represent a nondegenerate finite-area symmetry sphere.

Using the vacuum trace equation~(\ref{eq:X_prop}) to eliminate \(\hat X\), the normalized effective curvature-fluid scalars become:
\begin{align}
    &\bar{\mu} = \frac{\bar{R}}{6} + \frac{f(R)}{6\phi^{2}f'(R)}-\frac{f''(R)}{f'(R)}\bar{X}\bar{A}\bar{R}\phi^{2},\label{eq:mu_Rn}\\
    &\bar{p} = -\frac{5\bar{R}}{18} + \frac{f(R)}{18\phi^{2}f'(R)} - \frac{f''(R)}{3f'(R)}\bar{X}\bar{A}\bar{R}\phi^{2},\label{eq:p_Rn}\\
    &\bar{\Pi} = -\frac{2\bar{R}}{9} + \frac{4f(R)}{9\phi^{2}f'(R)}-\left(1+\frac{2}{3}\bar A\right)\frac{f''(R)}{f'(R)}\bar{X}\bar{R}\phi^{2}.\label{eq:Pi_Rn}
\end{align}
After Eqs.~\eqref{eq:mu_Rn}--\eqref{eq:Pi_Rn} are substituted into Eqs.~\eqref{eq: dAbar general f}--\eqref{eq: Xi general}, each specified \(\mathcal C^3\) function \(f(R)\) defines a local four-dimensional autonomous system for
\[
\left(\bar A,\bar R,\bar X,\phi\right)
\]
on every connected region of the chart where \(\phi Rf'f''\neq0\). The quantities \(\bar\mu,\bar p,\bar\Pi,\Xi,\bar K\), and \(\bar Z\) are then derived from these four variables. 

The power-law specialization below removes the explicit dependence of the normalized subsystem on $\phi$. The remaining equation $\phi_{,\rho}=\Xi\phi$ then reconstructs the dimensional scale along each normalized orbit.

\section{Qualitative analysis of the minimal vacuum system}\label{subsec:qualitative_reduced_system}
We now specialize to the signed real continuation of the power-law model,
\begin{equation}
\label{model_R_delta}
f(R)=R\left|\frac{R}{R_0}\right|^\delta .
\end{equation}
Here \(R_0>0\) supplies the curvature scale required for dimensional consistency and cancels from the vacuum autonomous system. On the positive-\(R\) branch this model coincides with the usual expression proportional to \(R^{1+\delta}\). The absolute value provides a real continuation to \(R<0\) for noninteger \(\delta\), which is essential because the Clifton--Barrow equilibrium identified below belongs to the negative-\(R\) sector for small positive \(\delta\). The model remains homogeneous under curvature rescalings, so it introduces no intrinsic scale into the reduced vacuum dynamics. Although the analysis is carried out for generic allowed values of \(\delta\), particular attention will be given to \(0<\delta\ll1\).

Away from $R=0$, one has
\begin{equation}
    \label{model_R_delta_deriv}
f'=(1+\delta)\left|\frac{R}{R_0}\right|^{\delta}, \quad\mbox{and}\quad f''=\frac{\delta(1+\delta)}{R}\left|\frac{R}{R_0}\right|^{\delta}.
\end{equation}
For \(0<\delta<1\) and \(R\neq0\), one has \(f'>0\) in both curvature-sign sectors, whereas \(f''\) has the sign of \(R\). In particular, \(f''<0\) on the negative-\(R\) branch containing \(P_{\rm C}\). At \(R=0\), however, \(f'(0)=0\) and \(f''\) is singular, so the Einstein-like equations obtained by dividing by \(f'\) do not extend to this boundary. Indeed, the undivided vacuum equations formally lose rank for \(R\equiv0\); consequently, scalar flatness does not by itself select the Schwarzschild geometry.

Substitution of Eqs.~\eqref{model_R_delta} and \eqref{model_R_delta_deriv} removes the explicit $\phi$ dependence from the normalized right-hand sides. Thus $\Xi$ becomes a function of $(\bar A,\bar R,\bar X)$, and $\phi_{,\rho}=\Xi\phi$ decouples as a reconstruction equation. The closed three-dimensional subsystem is

\begin{align}
&\bar A_{,\rho} = -\mathcal H(\bar A,\bar X)\bar A + b_0\bar R\bar A - b_1\bar R,&\label{eq:A_reduced_system}\\[1ex]
&\bar R_{,\rho} = \mathcal G(\bar A,\bar R,\bar X)\, \bar R,&\label{eq:R_reduced_system}\\
&\bar X_{,\rho} = - \mathcal H(\bar A,\bar X)\bar X +\frac{b_0}{\delta}(1+\delta\bar X)\bar R,&\label{eq:X_reduced_system}
\end{align}
where the auxiliary functions are
\begin{align}
\mathcal H(\bar A,\bar X) &\equiv \frac12+2\bar A(1+\delta\bar X)+2\delta\bar X,\label{eq:aux_H}\\
\mathcal G(\bar A,\bar R,\bar X) &\equiv 1-2\bar A(1+2\delta\bar X)+(1-2\delta)\bar X+2b_0\bar R,\label{eq:aux_G}
\end{align} 
with 
\begin{align}
    b_0=\frac{(1-\delta)}{3(1+\delta)},\qquad 
    b_1=\frac{(1+2\delta)}{6(1+\delta)}.
\end{align}
In particular, the Gaussian curvature is reconstructed via the constraint, Eq.~\eqref{eq: Kbar general}, and simplifies to
\begin{align}
\bar K = \frac14+\bar A +\delta(1+\bar A)\bar X +\frac{\delta}{2(1+\delta)}\bar R.\label{constraint1}
\end{align}
In terms of the normalized variables, the curvature fluid quantities are also simplified to
\begin{eqnarray}
\bar\mu &=& \frac{\delta+2}{6(1+\delta)}\bar R - \delta\bar A\bar X,\label{eq:muRn}\\
\bar{p} &=& -\frac{5\delta+4}{18(1+\delta)}\bar R - \frac{\delta}{3}\bar A\bar X,\label{eq:pRn}\\
\bar\Pi &=& \frac{2(1-\delta)}{9(1+\delta)}\bar R -\delta\left(1+\frac23\bar A\right)\bar X.\label{eq:piRn}
\end{eqnarray}
The generic reduced system assumes $\delta\neq0,-1$. The cases $\delta=0$ and $\delta=-1$ must be treated in the original undivided field equations, since the displayed reduced coefficients are undefined there. Other exceptional values, including $\delta=1$ and $\delta=-1/2$, require separate treatment wherever the fixed-point derivation divides by $b_0$ or $b_1$.

\subsection{Finite critical sets}\label{subsec:finite_critical_sets}

Finite equilibria satisfy $\bar A_{,\rho}=\bar R_{,\rho}=\bar X_{,\rho}=0$. Because $\bar R_{,\rho}=\mathcal G\bar R$, their calculation separates into $\bar R=0$ and $\mathcal G=0$. For $b_0\ne0$, the latter is
\begin{equation}
 \bar R=\bar R_*^{(2)} =-\frac{1-2\bar A(1+2\delta\bar X)+(1-2\delta)\bar X}{2b_0}.
\end{equation}
The polynomial extension preserves $\bar R=0$ and the sign of $\bar R$ on either side. Its zero-curvature plane is a formal boundary: $\bar X$ is undefined on an exact scalar-flat metric. The variable $\bar Z=X/\phi^3$ can record such a metric's curvature data when $\phi\ne0$, but does not remove the degeneracy caused by $f'(0)=0$.

By contrast, $\bar R=\bar R_*^{(2)}$ is only the nullcline $\bar R_{,\rho}=0$. It is not generally invariant and may be imposed when solving for equilibria, not when defining a two-dimensional dynamical subsystem. We consider these two algebraic branches in turn.

\subsubsection{Critical sets with vanishing Ricci scalar}
On the formal boundary $\bar R=0$, the reduced system becomes
\begin{align}
\bar A_{,\rho} &= -2\bar A \left[ \bar A(1+\delta\bar X) +\delta\bar X +\frac14 \right],\label{eq:A_red_R0}\\[1ex]
\bar X_{,\rho} &= -2\bar X \left[\bar A(1+\delta\bar X) +\delta\bar X +\frac14 \right].\label{eq:X_red_R0}
\end{align}
There is one isolated scalar-flat boundary point,
\begin{equation}
P_{\rm S}: \qquad (\bar A_{*},\bar R_{*},\bar X_{*})=(0,0,0). \label{eq: Ps point}
\end{equation}
Equation~\eqref{constraint1} gives the normalized value
\[
\bar K(P_{\rm S})=\frac14.
\]

This does not imply a nonzero physical Gaussian curvature at the asymptotic endpoint, since \(K=\phi^2\bar K\to0\) as \(r\to\infty\). The interpretation of this boundary requires the distinction emphasized by Nzioki et al.~\cite{Nzioki:2009av,Nzioki:2013lca}. When \(f(0)=0\), \(f\) is sufficiently differentiable at \(R=0\), and \(f'(0)\neq0\), the scalar-flat static spherically symmetric vacuum equations reduce to those of GR and uniquely yield the Schwarzschild solution \cite{Nzioki:2013lca}. The present signed power-law model instead satisfies \(f'(0)=0\) for positive \(\delta\), causing the Jebsen-Birkhoff generalization \cite{Nzioki:2013lca} to break down. As a result, the field equations lose rank at $R=0$, and the scalar-flat condition alone is no longer sufficient to uniquely select the Schwarzschild solution. Accordingly, we shall use a separately selected GR-like Ricci-flat extension, reconstructed in Sec.~\ref{sec:metric_reconstruction_fixed_points}. Within this extension, \(P_{\rm S}\) is the Minkowski endpoint approached by finite-mass Schwarzschild orbits, but the boundary point \(P_{\rm S}\) by itself does not uniquely determine the Schwarzschild geometry.

In addition, since the expressions between the brackets in Eqs.\ \eqref{eq:A_red_R0} and \eqref{eq:X_red_R0} are the same, there is a one-parameter critical curve whose coordinates are
\begin{equation}
\mathcal L_0(\sigma): \qquad (\bar A_*,\bar R_*,\bar X_*) = \left(-\frac{1+4\delta\sigma}{4(1+\delta\sigma)},0,\sigma \right), 
\label{eq:L0_reduced}
\end{equation}
with $\sigma\neq-1/\delta$. Along this curve, the Gaussian curvature vanishes according to Eq.\ (\ref{constraint1}). Thus, \(\mathcal L_0(\sigma)\) lies on the degenerate Gaussian-curvature boundary.

For nonstationary trajectories with $\bar A\neq0$, the common factor in Eqs.\ \eqref{eq:A_red_R0} and \eqref{eq:X_red_R0} yields the first integral
\begin{equation}
\frac{d}{d\rho}\left(\frac{\bar X}{\bar A}\right)=0, \quad\Longrightarrow\quad \bar X=c\bar A,
\label{eq:R0_first_integral}
\end{equation}
where $c$ is a constant. Thus, using the Gaussian curvature \eqref{constraint1}, the radial flow on each such line becomes
\begin{equation}
\bar A_{,\rho}=-2\bar K\bar A, \quad\mbox{with}\quad \bar K(\bar A)=\frac14+(1+\delta c)\bar A+\delta c\bar A^2
\label{eq:R0_line_flow}
\end{equation}
Hence, these lines, the invariant axes $\bar A$ and $ \bar X$, and their stationary points exhaust the flow on the formally extended $\bar R=0$ plane.

\subsubsection{Critical sets with nonvanishing Ricci scalar}
To determine equilibria in the nonzero-\(\bar R\) sector, we substitute the nullcline condition \(\bar R=\bar R_*^{(2)}\) into the remaining equilibrium equations, which gives
\begin{align}
   0 = &-\bar{A}^{2}-\bar{A}\left[\frac{1}{2}\left(2+\bar{X}+2\bar{X}\delta\right) + \frac{b_{1}}{b_{0}} \left(1 + 2\bar{X} \delta \right)\right]\nonumber\\
   &+\frac{b_{1}}{2b_{0}}\left(1+\bar{X}-2\bar{X}\delta\right),\nonumber\\
    0 = &-\frac{1}{2\delta}\left[1 + \bar{X} + \delta\left(1+2\delta\right)\bar{X}^{2} - 2\bar{A}\left(1+\delta\bar{X}\right)\right]. \label{eq: AX critical set nonvanishing Ricci}
\end{align}
Any solutions of these algebraic equations with $\bar R_*^{(2)}=0$ already belong to $\mathcal L_0$ and are excluded from the nonzero-curvature branch.

The remaining fixed-point equations can be simplified, yielding
\begin{equation}
0= -\bar{R}_{*}^{(2)}\left(b_1\bar X_*+\frac{b_0}{\delta}\bar A_*\right).
\end{equation}
Since \(\bar{R}_{*}^{(2)}\neq0\), one obtains
\begin{equation}
\bar A_* = -\frac{\delta b_1}{b_0}\bar X_*.
\end{equation}
Substitution into either remaining equation gives the factorized condition
\begin{equation}
(1+\bar X_*) \left[1-\delta+\delta(1+2\delta)\bar X_*\right]=0.
\end{equation}

For \(\delta\neq-1/2\), the second factor gives
\begin{equation}
P_{K0}: \quad (\bar A_*,\bar R_*,\bar X_*) = \left(\frac12,\frac{1-4\delta}{4\delta b_1}, -\frac{b_0}{2\delta b_1}\right). \label{eq: Pk0 point}
\end{equation}
The constraint gives $\bar K(P_{K0})=0$. Thus \(P_{K0}\) also belongs to the zero-Gaussian-curvature boundary. At \(\delta=1/4\), its Ricci coordinate also vanishes and
\[
P_{K0}\Big|_{\delta=\frac14}=\left(\frac12,0,-2\right) \in\mathcal L_0;
\]
it is then no longer a distinct nonzero Ricci scalar equilibrium.

The root \(\bar X_*=-1\) gives
\begin{equation}
P_{\rm C}: \quad (\bar A_*,\bar R_*,\bar X_*) = \left(\frac{b_1\,\delta}{b_0}, -\frac{\delta(4\delta^2-2\delta+1)}{6\,b_0^2\,(1+\delta)}, -1\right). \label{eq: Pc point}
\end{equation}
This corresponds dynamically to the scale-invariant fixed point identified with the Clifton--Barrow solution \cite{CliftonBarrow2005,Clifton:2006ug} (see its metric representation in Sec.~\ref{sec:metric_reconstruction_fixed_points}). Its normalized Gaussian curvature is
\begin{equation}
\bar K(P_{\rm C}) = \frac{(1-2\delta-2\delta^2)(4\delta^2-2\delta+1)}{4(1-\delta)^2}.
\end{equation}
Since
\begin{equation}
4\delta^2-2\delta+1 = 4\left(\delta-\frac14\right)^2+\frac34>0,
\end{equation}
the condition \(\bar K(P_{\rm C})>0\) restricts $\delta$ to
\begin{equation}
\frac{-1-\sqrt3}{2} < \delta < \frac{-1+\sqrt3}{2}.
\end{equation}
In the positive-\(\delta\) sector this reduces to
\begin{equation}
0<\delta<\frac{\sqrt3-1}{2}.
\end{equation}
For \(0<\delta<1/4\), one has \(\bar R(P_{K0})>0\), whereas \(\bar R(P_{\rm C})<0\). Hence these equilibria belong to different invariant \(\bar R\)-sign sectors, and no orbit remaining within the \(R\neq0\) reduced chart can connect them. The absolute-value prescription in Eq.~\eqref{model_R_delta} keeps the negative-\(R\) equilibrium real for noninteger \(\delta\) without changing the autonomous system within either fixed-sign sector.

\subsection{Outward-radial linear stability}\label{out_lin_stab}

We now linearize the flow about the finite critical sets. All source and sink classifications refer to increasing \(\rho\), equivalently, increasing areal radius. Therefore, stability here is described with respect to the radial asymptotics. The local classifications below use the standard linearization, invariant-manifold, and normal-hyperbolicity results of nonlinear dynamical-systems theory \cite{Wiggins2003}.

Let $\mathbf{y}=(\bar A,\bar R,\bar X)$ denote the state vector, and denote the right-hand sides of Eqs.~\eqref{eq:A_reduced_system}--\eqref{eq:X_reduced_system} by \(\mathbf{F}=(F_A,F_R,F_X)\). Using the auxiliary functions \eqref{eq:aux_H} and \eqref{eq:aux_G}, the Jacobian matrix is
\begin{widetext}
\begin{equation}
J(\mathbf{y}) = \frac{\partial\mathbf{F}}{\partial\mathbf{y}} =
\begin{pmatrix}
-\mathcal H-2\bar A(1+\delta\bar X)+b_0\bar R & b_0\bar A-b_1 & -2\delta\bar A(1+\bar A)\\[1ex]
-2\bar R(1+2\delta\bar X) & \mathcal G+2b_0\bar R & \bar R(1-2\delta-4\delta\bar A) \\[1ex]
-2\bar X(1+\delta\bar X) & \dfrac{b_0}{\delta}(1+\delta\bar X) & -\mathcal H-2\delta(1+\bar A)\bar X+b_0\bar R
\end{pmatrix},
\label{eq:Jacobian_reduced}
\end{equation}
\end{widetext}
Since the reduced system is three-dimensional, each critical set has three eigenvalues.

\subsubsection{The Schwarzschild endpoint} 
\begin{figure}[htb]
    \centering
    \includegraphics[scale=0.4]{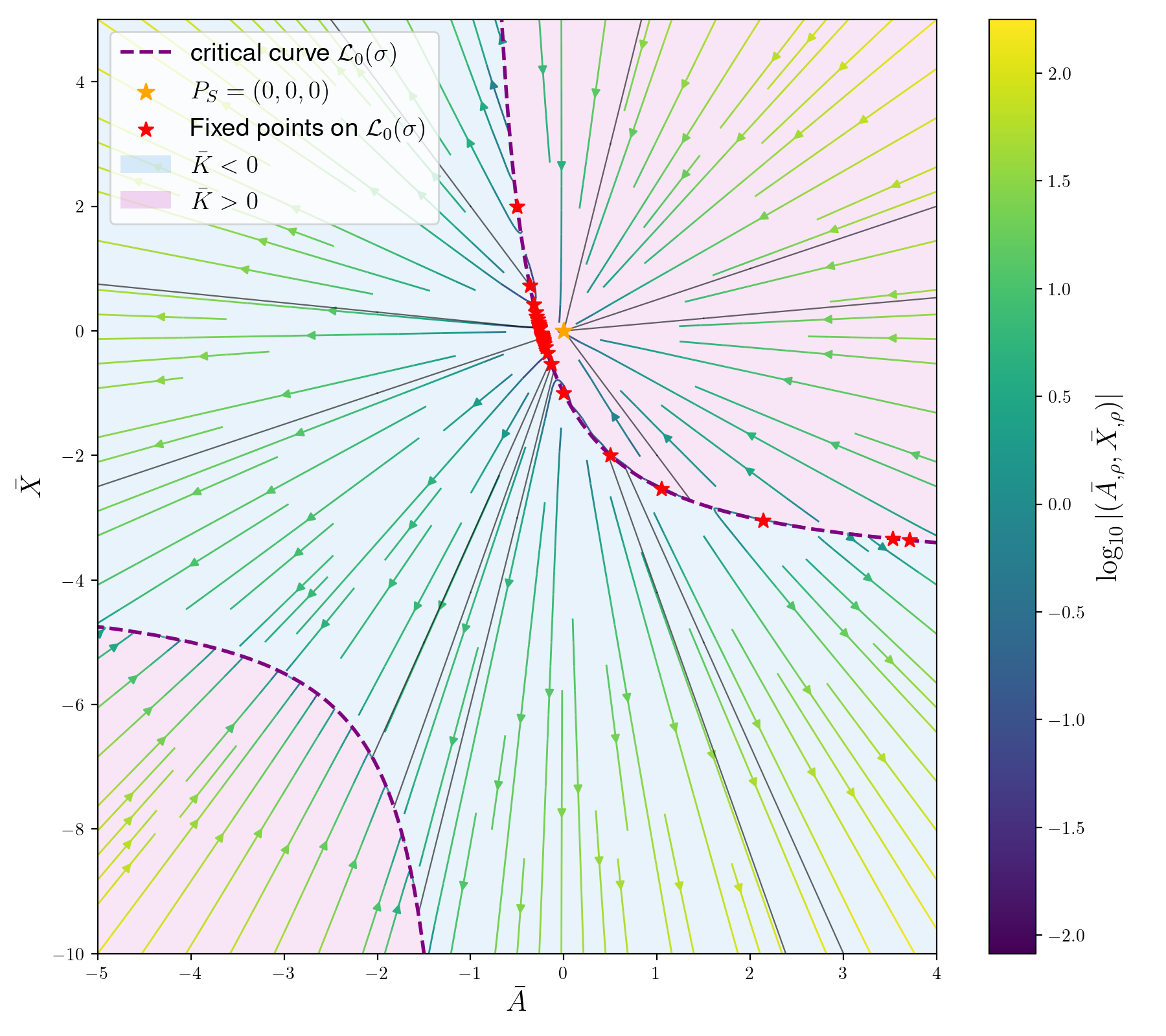}
    \caption{Phase space for the reduced system of Eqs. \eqref{eq:A_red_R0} and \eqref{eq:X_red_R0} on the invariant plane $\bar{R}=0$ of the polynomially extended system. We choose $\delta=1/4$; $\bar X$ labels formal limiting directions and is undefined on an exact scalar-flat metric. The regions of positive and negative normalized Gaussian curvature are shaded in pink and blue, respectively. The critical curve $\mathcal{L}_{0}(\sigma)$ has zero Gaussian curvature. The colorbar shows the magnitude of the velocity vector of the flow in $\log_{10}$ scale.}
    \label{fig:2d phase space reduced sys R0}
\end{figure}

At the critical point $P_{\rm S}$, Eq.~\eqref{eq: Ps point}, the spectrum is
\begin{equation}
\operatorname{spec}J(P_{\rm S}) = \left\{ 1,-\frac12,-\frac12 \right\}.
\label{eq:eigenvalues_PS}
\end{equation}
The positive eigenvalue is normal to $\bar R=0$, whereas the two negative eigenvalues are tangent to that plane. Thus $P_{\rm S}$ is a hyperbolic saddle of the polynomially extended three-dimensional system and a local attractor within its formal scalar-flat plane (Fig.~\ref{fig:2d phase space reduced sys R0}). This classification is independent of $\delta$ wherever the generic reduced system is defined. The plane is not a Ricci-flat sector: only a separately selected subfamily reconstructs Schwarzschild. Nor is $P_{\rm S}$ a global attractor on the plane, since $\mathcal L_0$ separates components with different radial endpoints.

\subsubsection{The critical curve \texorpdfstring{$\mathcal L_0(\sigma)$}{L0(sigma)}}

At the critical curve $\mathcal L_0(\sigma)$, Eq.~\eqref{eq:L0_reduced}, the eigenvalues are
\begin{align}
\lambda_{\parallel}&=0,\label{eq:L0_lambda_parallel}\\
\lambda_{\perp}&=\frac{4\delta^2\sigma^2+2\delta\sigma+1}{2(1+\delta\sigma)},\label{eq:L0_lambda_perp}\\
\lambda_{\bar R}&=\frac{2\delta(1+2\delta)\sigma^2+2(1+2\delta)\sigma+3}{2(1+\delta\sigma)}.\label{eq:L0_lambda_R}
\end{align}
Here \(\lambda_{\parallel}=0\) is tangent to the critical curve, \(\lambda_{\perp}\) is transverse to the curve but tangent to the scalar-flat surface, and \(\lambda_{\bar R}\) is normal to \(\bar R=0\).

Although every point of \(\mathcal L_0\) is non-hyperbolic when regarded as an isolated equilibrium, the curve is normally hyperbolic wherever both transverse eigenvalues are nonzero. In particular, the sign of \(\lambda_{\perp}\) is determined entirely by its denominator \(1+\delta\sigma\).

The $\bar R$--normal eigenvalue $\lambda_{\bar R}$ vanishes at
\begin{equation}
\sigma_{\pm} = \frac{-(1+2\delta)\pm\sqrt{(1+2\delta)(1-4\delta)}}{2\delta(1+2\delta)},
\label{eq:sigma_pm}
\end{equation}
for $\sigma_{\pm}\in\mathbb{R}$. In these cases, \(\mathcal L_0\) loses normal hyperbolicity and a center-manifold analysis would be required to determine the nonlinear flow in the two-dimensional center subspace.

For \(0<\delta<1/4\), one has
\begin{equation}
-\frac1\delta<\sigma_-<\sigma_+.
\end{equation}
On the branch \(1+\delta\sigma>0\), the curve is normally repelling for
\begin{equation}
-\frac1\delta<\sigma<\sigma_- \qquad\text{or}\qquad \sigma>\sigma_+,
\end{equation}
and normally saddle-like for
\begin{equation}
\sigma_-<\sigma<\sigma_+.
\end{equation}
The disconnected branch \(\sigma<-1/\delta\) is normally attracting. Since the entire curve satisfies \(\bar K=0\), none of these sectors represents a regular spherical region with \(\bar K>0\).

Equation~\eqref{eq:R0_line_flow} makes the disconnected positive-$\bar K$ component explicit. At $\bar A=-1$ one has $\bar K(-1)=-3/4$ for every $\delta c$, so a trajectory with $\bar A<-1$ and $\bar K>0$ cannot reach $P_{\rm S}$ while remaining in the physical spherical sector. Such trajectories occur for $\delta c>0$: outward radial evolution approaches the $\bar K=0$ curve $\mathcal L_0$, whereas inward evolution has $\bar A\to-\infty$. Their distinct metric endpoints are reconstructed in Sec.~\ref{subsec:other_fixed_points_metric_current}.

\subsubsection{Zero-Gaussian-curvature point \texorpdfstring{$P_{K0}$}{PK0}}

\begin{figure}[htb]
    \hspace*{-1cm}\includegraphics[scale=0.43]{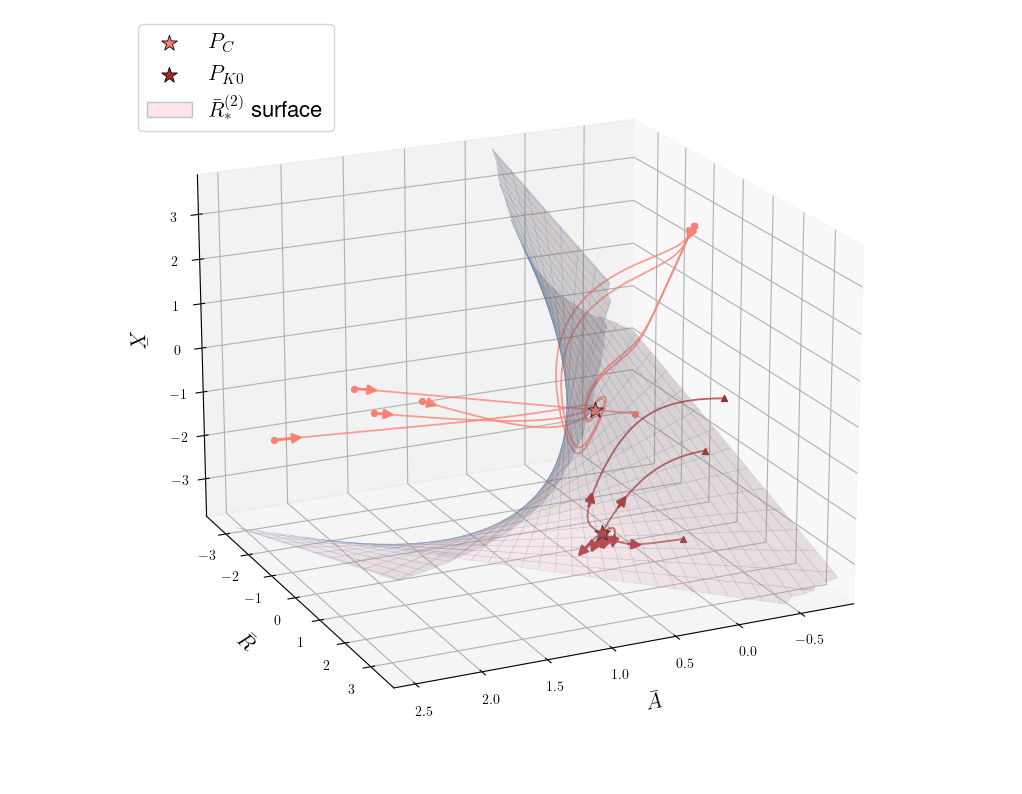}
    \caption{Three-dimensional phase portrait of the full system \eqref{eq:A_reduced_system}--\eqref{eq:X_reduced_system} for \(\delta=1/5\) in the nonzero-\(\bar R\) chart. The shaded surface shows the nullcline \(\bar R=\bar R_*^{(2)}\). Since \(\bar R=\bar R_*^{(2)}\) is not invariant, the trajectories were obtained from the full three-dimensional system and may cross this surface. The stars mark \(P_{\rm C}\) and \(P_{K0}\), the dots indicate the initial conditions, and the arrows show the direction of increasing \(\rho\).}
    \label{fig:3d phase space reduced sys Rneq0}
\end{figure}

At the critical point \(P_{K0}\), Eq.~\eqref{eq: Pk0 point}, the spectrum is
\begin{equation}
\operatorname{spec}J(P_{K0}) = \left\{\frac{1-4\delta}{2\delta},\frac{1-4\delta}{2\delta},-\frac{2\delta^2+2\delta-1}{\delta(1+2\delta)}\right\}.
\label{eq:eigenvalues_PK0}
\end{equation}
Hence, in the positive-\(\delta\) sector, one has that: \(P_{K0}\) is a source for \(0<\delta<1/4\), joining \(\mathcal L_0\) at \(\delta=1/4\); it is a saddle for $\frac14<\delta<\delta_{\rm K}$ with $\delta_{\rm K}\equiv\frac{\sqrt3-1}{2}$, and it coalesces with \(P_{\rm C}\) at \(\delta=\delta_{\rm K}\), with one zero eigenvalue; for $\delta_{\rm K}<\delta<1$, it is a sink on the boundary $\bar K=0$. In particular, \(P_{K0}\) is a hyperbolic source for \(0<\delta\ll1\), see Fig.~\ref{fig:3d phase space reduced sys Rneq0}. Its stability is relevant to the global organization of the phase portrait, but it does not describe a regular spherical solution because \(\bar K(P_{K0})=0\). Furthermore, at \(\delta=1/4\) and \(\delta=\delta_{\rm K}\), the vanishing eigenvalues prevent a stability classification by linearization alone; a center-manifold or bifurcation analysis is required.

\subsubsection{The Clifton--Barrow point \texorpdfstring{$P_{\rm C}$}{PC}}

At the critical point \(P_{\rm C}\), with coordinates given by Eq.~\eqref{eq: Pc point}, the eigenvalues are
\begin{equation}
\label{eq:PC_eigenvalues}
\lambda_1=\frac{N_\delta}{2(\delta-1)}, \qquad \lambda_{\pm}=\frac{N_\delta\pm\sqrt{\Delta_{\rm C}}}{4(\delta-1)},
\end{equation}
where $N_\delta\equiv4\delta^2-2\delta+1$ and $\Delta_{\rm C}\equiv N_{\delta}\left(2 \delta  \left(10 \delta +7\right)-7\right)$. Since $N_\delta>0$, one has \(\lambda_1<0\) whenever \(\delta<1\). Furthermore,
\begin{equation}
\lambda_+\lambda_- = -\frac{N_\delta(2\delta^2+2\delta-1)}{2(\delta-1)^2}.
\label{eq:PC_pair_product}
\end{equation}
It follows that all three eigenvalues have negative real parts throughout the regular spherical interval
\begin{equation}
0<\delta<\delta_{\rm K}.
\end{equation}

The transition from spiral to node is controlled by $\Delta_{\rm C}$ whose relevant positive root is
\begin{equation}
\delta_{\rm sp} = \frac{-7+3\sqrt{21}}{20}\simeq0.337386.
\end{equation}
Consequently:

\begin{itemize}
\item for \(0<\delta<\delta_{\rm sp}\), \(P_{\rm C}\) is a stable focus;
\item at \(\delta=\delta_{\rm sp}\), the complex pair coalesces into a repeated negative real eigenvalue;
\item for \(\delta_{\rm sp}<\delta<\delta_{\rm K}\), \(P_{\rm C}\) is a stable node;
\item at \(\delta=\delta_{\rm K}\), one eigenvalue vanishes and \(P_{\rm C}\) coalesces with \(P_{K0}\);
\item for \(\delta_{\rm K}<\delta<1\), \(P_{\rm C}\) is a saddle and lies outside the regular sector \(\bar K>0\).
\end{itemize}
Thus, \(P_{\rm C}\) is an outward-radial sink throughout the interval $0<\delta<\delta_{\rm K}$, in which \(\bar K(P_C)>0\), as illustrated in Fig.~\ref{fig:3d phase space reduced sys Rneq0}. An important qualification concerns the physical stability of this equilibrium. The Clifton–Barrow solution has $R<0$ throughout this parameter interval and consequently satisfies $f'>0$ but $f''<0$. The latter violates the usual Dolgov–Kawasaki viability condition for metric $f(R)$ gravity. Indeed, the commonly used scalaron mass expression $m_{\mathrm{eff}}^2=(f'-Rf'')/(3f'')$ gives $m_{\mathrm{eff}}^2=(1-\delta)R/(3\delta)<0$ for $0<\delta<\delta_K$. This indicates a potential scalar instability, although the Clifton–Barrow background has a nonconstant Ricci scalar and a definitive assessment requires a time-dependent perturbation analysis of the full background. The radial attraction established here must therefore be understood as a property of the static spatial dynamical system, rather than as evidence of dynamical stability under physical perturbations.

\subsubsection{Small positive deviations from GR}

For \(0<\delta\ll1\), the Schwarzschild/Minkowski endpoint \(P_{\rm S}\) is a saddle of the full reduced system and is attracting only within the invariant scalar-flat boundary, even though it is not a global attractor in the subspace (see Fig.~\ref{fig:2d phase space reduced sys R0}). For finite \(\sigma\), the curve \(\mathcal L_0\) is normally repelling for \(\sigma>\sigma_+\simeq-3/2\) and normally saddle-like for \(\sigma<\sigma_+\). The point \(P_{K0}\) is a source on the degenerate boundary \(\bar K=0\), whereas the Clifton--Barrow point \(P_{\rm C}\) is a spiral sink. Therefore, \(P_{\rm C}\) is the only regular finite attractor of the full three-dimensional system in the small-positive-\(\delta\) regime.

\subsection{Radial compactification and asymptotic critical sets}
\label{subsec:compactification_infinity_reduced}

In order to study the asymptotic stability of the vacuum system \eqref{eq:A_reduced_system}--\eqref{eq:X_reduced_system}, we employ the polar representation of the Poincar\'e compactification. With this technique, it is possible to determine the leading critical sets and their normal flow at infinity (see Ref.~\cite{Matsue2018} for details).

Since the vector field is polynomial of degree three, it is useful to decompose it into homogeneous components,
\begin{equation}
\mathbf{F} = \mathbf{F}_3+\mathbf{F}_2+\mathbf{F}_1.
\label{eq:F_homogeneous_decomposition}
\end{equation}
The cubic part is
\begin{equation}
\mathbf{F}_3 = -2\delta\bar A\bar X
\begin{pmatrix}
\bar A\\
2\bar R\\
\bar X
\end{pmatrix},
\label{eq:F3}
\end{equation}
whereas the quadratic part is
\begin{equation}
\mathbf{F}_2 =
\begin{pmatrix}
-2\bar A^2-2\delta\bar A\bar X+b_0\bar A\bar R\\[1ex]
-2\bar A\bar R+(1-2\delta)\bar R\bar X +2b_0\bar R^2\\[1ex]
-2\bar A\bar X-2\delta\bar X^2+b_0\bar R\bar X
\end{pmatrix}.
\label{eq:F2}
\end{equation}
Finally, the linear contribution reads
\begin{equation}
\mathbf{F}_1 =
\begin{pmatrix}
-\dfrac12\bar A-b_1\bar R\\[1ex]
\bar R\\[1ex]
-\dfrac12\bar X+\dfrac{b_0}{\delta}\bar R
\end{pmatrix}.
\label{eq:F1}
\end{equation}

Asymptotically, $\mathbf{F}_3$ plays the most important role, while the other components come into play when zero modes require additional resolution.

\subsubsection{The Poincaré sphere}

Let us introduce the radial compactification variables
\begin{equation}
\mathbf{u} = (u_A,u_R,u_X)=z\,\mathbf{y}, \quad\mbox{with}\quad u_A^2+u_R^2+u_X^2=1,
\label{eq:Poincare_variables}
\end{equation}
where \(z=\|\mathbf{y}\|^{-1}\), so that \(z=0\) represents infinity. With the desingularized radial parameter $\tau$ defined as
\begin{equation}
\frac{d\tau}{d\rho}=z^{-2},
\label{eq:Poincare_time}
\end{equation}
the exact compactified system can be expressed as 
\begin{align}
\mathbf{u}_{,\tau} ={}& \left( I-\mathbf{u}\mathbf{u}^{\mathsf T} \right) \left[ \mathbf{F}_3(\mathbf{u}) +z\mathbf{F}_2(\mathbf{u}) +z^2\mathbf{F}_1(\mathbf{u}) \right], \label{eq:compactified_u}\\
z_{,\tau} ={}& -z\left( \Sigma_3+z\Sigma_2+z^2\Sigma_1 \right), \label{eq:compactified_z}
\end{align}
where
\begin{equation}
\Sigma_k \equiv \mathbf{u}\cdot\mathbf{F}_k(\mathbf{u}).
\end{equation}
Thus, the lower-degree terms do not modify the flow obtained by setting \(z=0\), although they become essential whenever the cubic contribution is degenerate. In particular, for Eq.~\eqref{eq:F3}, one obtains
\begin{equation}
\Sigma_3 = -2\delta u_Au_X \left(u_A^2+2u_R^2+u_X^2\right).
\end{equation}

A point at infinity is critical when \(\mathbf{F}_3(\mathbf{u})\) is parallel to \(\mathbf{u}\). The complete cubic critical set of Eqs.~\eqref{eq:compactified_u}--\eqref{eq:compactified_z} is therefore
\begin{equation}
\mathscr I_\infty = \{u_R=0\} \cup \{u_A=0\} \cup \{u_X=0\}
\label{eq:I_infinity}
\end{equation}
where $z=0$ for each subset. Hence, infinity does not consist of isolated critical points, but of three intersecting great circles.

The normally hyperbolic part is
\begin{equation}
\mathscr C_{AX}: \qquad z=0, \quad u_R=0, \quad u_Au_X\neq0, \quad u_A^2+u_X^2=1.
\end{equation}
At each point of this set, the spectrum contains one zero eigenvalue tangent to the critical circle and the two transverse eigenvalues
\begin{equation}
\lambda_R=-2\delta u_Au_X, \qquad \lambda_z=2\delta u_Au_X.
\label{eq:CAX_eigenvalues}
\end{equation}
These eigenvalues always have opposite signs. Therefore, \(\mathscr C_{AX}\) is normally saddle-like and contains neither sinks nor sources of the full compactified system.

The Gaussian-curvature constraint \eqref{constraint1} gives, along this circle,
\begin{equation}
\bar K = \frac{\delta u_Au_X}{z^2} +O(z^{-1}).
\end{equation}
Consequently, for \(\delta>0\), the portion of \(\mathscr C_{AX}\) lying in the regular spherical sector \(\bar K>0\) has \(u_Au_X>0\). On this physical portion, \(\lambda_z>0\), so the boundary is repelling in the compactifying normal direction $z$.

\subsubsection{Degenerate projective directions}

On the circles $u_A=0$ and $u_X=0$, the cubic field vanishes. Nevertheless, a bounded original variable can multiply diverging variables and contribute at leading order. We therefore retain the full reduced vector field in the secondary charts below. Their classification is stated for $0<\delta<1$.

First consider the chart
\begin{equation}
 \bar{A}=\frac{\epsilon}{\zeta},\qquad  \bar{R}=\frac{v}{\zeta},\qquad \bar{X}=\chi,\qquad \frac{d\eta}{d\rho}=\zeta^{-1},
 \label{eq:purple_A_chart}
\end{equation}
with $\zeta>0$ and $\epsilon=\pm1$. Here $\chi$ is finite as $\zeta\to0$. Direct substitution into the full reduced system gives
\begin{align}
 \zeta_{,\eta} &=\zeta\left[2\epsilon(1+\delta\chi)-b_0v\right] +\zeta^2\left(2\delta\chi+\frac12+\epsilon b_1v\right),\label{eq:purple_A_zeta}\\
 v_{,\eta} &=v(b_0v-2\delta\epsilon\chi) +\zeta v\left(\chi+\frac32+\epsilon b_1v\right),\label{eq:purple_A_v}\\
 \chi_{,\eta} &=(1+\delta\chi)\left(-2\epsilon\chi+\frac{b_0}{\delta}v\right) -\zeta\chi\left(2\delta\chi+\frac12\right).\label{eq:purple_A_chi}
\end{align}
These equations extend regularly to $\zeta=0$. Their boundary equilibria consist of the line
\begin{equation}
 \mathcal E_A^\epsilon(\chi):\qquad \zeta=0,\qquad v=\frac{2\delta\epsilon}{b_0}\chi, \qquad \chi\in\mathbb R,
 \label{eq:purple_EA}
\end{equation}
and the isolated equilibrium
\begin{equation}
 \mathcal D_A^\epsilon:\qquad (\zeta,v,\chi)=(0,0,-1/\delta).
 \label{eq:purple_DA}
\end{equation}
The spectra of the full chart Jacobian are
\begin{align}
 \operatorname{spec}J(\mathcal E_A^\epsilon)  &=\{0,-2\epsilon,2\epsilon\}, \label{eq:purple_EA_spec}\\
 \operatorname{spec}J(\mathcal D_A^\epsilon)  &=\{0,2\epsilon,2\epsilon\}.  \label{eq:purple_DA_spec}
\end{align}
Thus $\mathcal E_A^\epsilon$ is normally saddle-like. Its zero eigenvalue is tangent to the critical line. Both $\mathcal E_A^\epsilon(0)$ and $\mathcal D_A^\epsilon$ project onto $Q_A^\epsilon=[\epsilon:0:0]$; the projective ray alone therefore does not specify a unique secondary equilibrium. The zero mode at $\mathcal D_A^\epsilon$ remains unresolved.

The complementary chart with bounded $\bar{A}$ is
\begin{equation}
 \bar{X}=\frac{\epsilon}{\zeta},\qquad \bar{R}=\frac{v}{\zeta},\qquad \bar{A}=\alpha,\qquad \frac{d\eta}{d\rho}=\zeta^{-1}.
 \label{eq:purple_X_chart}
\end{equation}
The exact transformed equations are
\begin{align}
\zeta_{,\eta} &=\zeta\left[2\delta\epsilon(\alpha+1)-b_0v\right] +\zeta^2\left(2\alpha+\frac12-\frac{\epsilon b_0}{\delta}v\right), \label{eq:purple_X_zeta}\\
v_{,\eta} &=v\left[\epsilon(1-2\delta\alpha)+b_0v\right] +\zeta v\left(\frac32-\frac{\epsilon b_0}{\delta}v\right), \label{eq:purple_X_v}\\
\alpha_{,\eta} &=-2\delta\epsilon\alpha(\alpha+1)+(b_0\alpha-b_1)v -\zeta\alpha\left(2\alpha+\frac12\right). \label{eq:purple_X_alpha}
\end{align}
At $\zeta=0$, the equilibria, in the order $(\zeta,v,\alpha)$, are
\begin{align}
 \mathcal P_X^\epsilon&=(0,0,0),& \mathcal D_X^\epsilon&=(0,0,-1),\nonumber\\
 \mathcal P_{RX}^\epsilon&= \left(0,-3(1+\delta)\epsilon,\frac12\right).
 \label{eq:equil_purple_PRX}
\end{align}
Their full-chart spectra are
\begin{align}
 \operatorname{spec}J(\mathcal P_X^\epsilon) &=\{-2\delta\epsilon,\epsilon,2\delta\epsilon\}, \label{eq:purple_PX}\\
 \operatorname{spec}J(\mathcal D_X^\epsilon) &=\{0,2\delta\epsilon,(1+2\delta)\epsilon\}, \label{eq:purple_DX}\\
 \operatorname{spec}J(\mathcal P_{RX}^\epsilon) &=\{-\epsilon,-(1+2\delta)\epsilon,(1+2\delta)\epsilon\}. \label{eq:purple_PRX}
\end{align}

The first two equilibria both project onto $Q_X^\epsilon=[0:0:\epsilon]$. The mixed equilibrium instead projects onto
\begin{equation}
 Q_{RX}^\epsilon=[0:-3(1+\delta)\epsilon:\epsilon].
 \label{eq:purple_mixed_ray}
\end{equation}
These lifts depend on the full transformed system; a truncation to $\mathbf F_2$ does not retain all terms of leading order in these charts.

For $0<\delta<1$, $\mathcal P_X^\epsilon$ and $\mathcal P_{RX}^\epsilon$ are hyperbolic saddles, whereas $\mathcal D_X^\epsilon$ requires a nonlinear analysis of its zero mode.

These are spectra of regular secondary charts of the full system, with increasing $\eta$ having the same orientation as increasing $\rho$. They are not the spectra in the original cubic time scale. They describe only the divergence regimes retained by the stated charts, and every geometric interpretation must also impose $\bar{K}>0$. Other subordinate divergence rates may require further blow-ups. The pure curvature direction $Q_R^\epsilon=[0:\epsilon:0]$, in particular, requires the weighted resolution developed next.

\subsubsection{Weighted resolution of the \texorpdfstring{\(\bar R\)}{R}-dominated direction}

Since the sign of \(\bar R\) is invariant, introduce $\epsilon\equiv\operatorname{sgn}\bar R=\pm1$ and the weighted variables
\begin{equation}
\bar R=\frac{\epsilon}{w^2}, \qquad \bar A=\frac{U}{w}, \qquad \bar X=\frac{V}{w}, \qquad w=|\bar R|^{-1/2}.
\label{eq:R_weighted_chart}
\end{equation}
All finite $(U,V)$ at $w=0$ project onto $Q_R^\epsilon$. These weights balance $\bar R^2$ and $\bar A\bar R\bar X$ at order $w^{-4}$ in the curvature equation, while the leading transverse terms $\bar A^2\bar X$ and $\bar A\bar X^2$ are of order $w^{-3}$.

Using the desingularized parameter
\begin{equation}
\frac{d\tau_R}{d\rho}=w^{-2}
\end{equation}
the full reduced system becomes
\begin{align}
w_{,\tau_R} ={}& w\left[ 2\delta UV-\epsilon b_0 +w\left( U-\frac{1-2\delta}{2}V \right) -\frac12w^2 \right], \label{eq:w_weighted}\\
U_{,\tau_R} ={}& w\left[ -U^2 -\left(\delta+\frac12\right)UV -\epsilon b_1 -Uw \right], \label{eq:U_weighted}\\
V_{,\tau_R} ={}& w\left[ -UV -\left(\delta+\frac12\right)V^2 +\epsilon\frac{b_0}{\delta} -Vw \right].\label{eq:V_weighted}
\end{align}
The \(\bar R\)-dominated boundary is therefore the two-dimensional critical sheet
\begin{equation}
\mathscr Q_R^\epsilon: \qquad w=0, \qquad (U,V)\in\mathbb R^2.
\label{eq:QR_sheet}
\end{equation}
Its eigenvalues are
\begin{equation}
\operatorname{spec} J\bigl(\mathscr Q_R^\epsilon\bigr) = \left\{ 0,\, 0,\, \lambda_w \right\}, \qquad \lambda_w = 2\delta UV-\epsilon b_0.
\label{eq:QR_weighted_spectrum}
\end{equation}
Thus, \(\mathscr Q_R^\epsilon\) is normally hyperbolic where \(\lambda_w\neq0\). It is normally attracting for \(\lambda_w<0\) and normally repelling when the inequality is reversed. It is not a hyperbolic sink because the \(U\)- and \(V\)-directions remain neutral at linear order.

The Gaussian curvature in this chart is
\begin{equation}
\bar K = \frac{1}{w^2} \left[ \delta UV +\epsilon\frac{\delta}{2(1+\delta)} \right] + \frac{U+\delta V}{w} +\frac14.
\label{eq:K_weighted}
\end{equation}
Consequently,
\begin{equation}
\frac{\bar K}{\bar R} \longrightarrow \frac{\delta}{2(1+\delta)} +\epsilon\delta UV.
\label{eq:K_over_R_weighted}
\end{equation}
This equation shows that the projective direction \(Q_R^\epsilon\) does not determine a unique limit of \(\bar K/\bar R\). The value
\[
\frac{\bar K}{\bar R}\longrightarrow \frac{\delta}{2(1+\delta)}
\]
is recovered only on the special subdirection \(UV=0\), or under the condition $\bar A\bar X=o(|\bar R|)$, equivalently $\bar A\bar X/\bar R\to0$. A big-$O$ bound alone permits a nonzero limiting contribution and is insufficient.

For $0<\delta<1$, $b_0>0$. On the positive-$\bar R$ sheet, normal attraction and a strictly positive leading coefficient of $\bar K$ occur simultaneously when
\begin{equation}
 -\frac{1}{2(1+\delta)}<UV<\frac{b_0}{2\delta}.
 \label{eq:physical_QR_plus}
\end{equation}
This gives a normally attracting family at large normalized curvature. At the lower endpoint, the $w^{-2}$ coefficient of $\bar K$ vanishes, so physical admissibility must instead be decided from subleading terms. At the upper endpoint, normal hyperbolicity is lost. Large physical $|R|$ does not follow without reconstructing $\phi$.

On the negative-$\bar R$ sheet, normal attraction requires $UV<-b_0/(2\delta)$, whereas a nonnegative leading coefficient of $\bar K$ requires $UV\geq1/[2(1+\delta)]$. These ranges are disjoint. Thus this sheet has no normally attracting part compatible with the regular spherical sector.

The compactification therefore supplies a local classification of the balances resolved by these charts. The physical part of $\mathscr C_{AX}$ repels in the compactifying normal direction, the secondary charts contain saddles and unresolved zero modes, and $\mathscr Q_R^+$ contains a normally attracting physical subset for $0<\delta<1$. The last set is separated from the negative-$\bar R$ Clifton--Barrow equilibrium by the invariant boundary $\bar R=0$. None of these projective data alone determines a spacetime metric. The next section reconstructs the radial scale and metric functions needed to interpret them geometrically.

\section{Metric reconstruction and interpretation of the fixed points}
\label{sec:metric_reconstruction_fixed_points}

The qualitative analysis above was performed using covariantly defined normalized variables. In this section we connect those variables with the standard metric representation of static, spherically symmetric spacetimes. This provides a geometrical interpretation of the fixed points found in the reduced system and clarifies which of them correspond to regular vacuum metrics.

\subsection{Dictionary between covariant and metric variables}
\label{subsec:covariant_metric_dictionary}

Consider the areal-radius form of a static, spherically symmetric metric,
\begin{equation}
ds^2 = -e^{2\nu(r)}dt^2 + e^{2\lambda(r)}dr^2 + r^2d\Omega^2 .
\label{eq:areal_radius_metric}
\end{equation}
The preferred radial vector is
\begin{equation}
e^a\partial_a=e^{-\lambda}\frac{d}{dr},
\end{equation}
so that, for any scalar \(S\),
\begin{equation}
\hat S=e^{-\lambda}S_{,r}.
\end{equation}
The relation with the radial variable $\rho$ comes directly from Eqs.~\eqref{eq:hat_rho_der_rel}--\eqref{eq:rho_r_rel}. 

The sheet expansion and the acceleration scalar are
\begin{equation}
\phi=\frac{2e^{-\lambda}}{r}, \qquad A=e^{-\lambda}\nu_{,r}.
\end{equation}
Thus, the normalized acceleration is related to the metric functions by
\begin{equation}
\bar A = \frac{A}{\phi} = \frac{r\nu_{,r}}{2} = \frac14\frac{d\ln |g_{tt}|}{d\ln r},
\label{eq:Abar_metric_dictionary}
\end{equation}
and the normalized Gaussian curvature is
\begin{equation}
\bar K = \frac{K}{\phi^2} = \frac{e^{2\lambda}}{4} = \frac{g_{rr}}{4},
\label{eq:Kbar_metric_dictionary}
\end{equation}
where \(K=r^{-2}\) is the Gaussian curvature of a symmetry two-sphere. 

Since $g^{rr}=g^{ab}\nabla_a r\nabla_b r$ in this metric dictionary, radial null vectors have sheet expansions proportional to $\phi$. Thus, $\phi=0$ for some $r$ is a candidate marginal sphere: a regular Killing horizon also requires $\nu\to-\infty$ and a regular extension. If $\nu$ remains finite, the same zero can describe a smooth minimal sphere. Event-horizon status requires the global causal extension.

On an exact scalar-flat configuration, $R=X=0$, so $\bar X$ is undefined. The direct definition $\bar Z=X/\phi^3$ remains meaningful where $\phi\ne0$ and gives $\bar Z=0$. This records the curvature data without assigning a value to $X/R$, but does not regularize the field equations divided by $f'$. Scalar-flat reconstructions below are therefore checked against the undivided equations. In the regular nonzero-$R$ chart, the curvature-fluid scalars follow algebraically from Eqs.~\eqref{eq:muRn}--\eqref{eq:piRn}.

Let \((\bar A_*,\bar R_*,\bar X_*)\) be a finite equilibrium, and let \(\bar K_*\) be reconstructed from the algebraic constraint \eqref{constraint1}. Eqs.~\eqref{eq:Abar_metric_dictionary} and \eqref{eq:Kbar_metric_dictionary} imply the local power-law metric
\begin{equation}
    \label{eq:power_law_metric_reconstruction_current}
ds^2 =-C_t\left(\frac{r}{r_0}\right)^{4\bar A_*}dt^2 +4\bar K_*dr^2+r^2d\Omega^2, \qquad C_t>0.
\end{equation}
Therefore, every physically reconstructible finite equilibrium corresponds, on this coordinate patch, to a self-similar static spherical geometry. Fixed points with \(\bar K_*=0\) are degenerate boundaries of the normalized phase space, whereas points with \(\bar K_*<0\) do not represent real Lorentzian metrics of the form considered here.

For the metric \eqref{eq:power_law_metric_reconstruction_current}, the normalized Ricci scalar is
\begin{equation}
\bar R_* = 2\bar K_* -2\bar A_*^2 -\bar A_* -\frac12 .
\label{eq:Rbar_metric_AK_relation}
\end{equation}
For a nondegenerate equilibrium, $\bar K_*>0$ and $\bar K_{,\rho}=0$ imply $\Xi=-1/2$. Hence $\phi\propto e^{-\rho/2}\propto r^{-1}$. If also $\bar R_*\ne0$, then $R=\bar R_*\phi^2\propto r^{-2}$ and
\begin{equation}
 \bar X_* = \frac{d\ln|R|}{d\rho}=-1.
 \label{eq:Xbar_minus_one_selfsimilar}
\end{equation}
Thus every nonzero-curvature equilibrium with $\bar K_*>0$ lies on the branch $\bar X=-1$.

\subsection{A radial first integral}
\label{subsec:radial_first_integral}

We first derive a radial first integral for the regular nonzero-\(R\) system. Define
\begin{equation}
 S=b_0\bar A+\delta b_1\bar X, \qquad \mathcal P=\frac{re^\nu f'}{\sqrt{\bar K}}.
 \label{eq:charge_factors}
\end{equation}
The reduced equations and reconstruction identities give
\begin{align}
 S_{,\rho}&=(b_0\bar R-\mathcal H)S,\nonumber\\
 \Xi&=-\frac12+\bar A+2\delta\bar A\bar X +\delta\bar X-b_0\bar R,\nonumber\\
 \frac{\mathcal P_{,\rho}}{\mathcal P} &=\frac12+\bar A+\delta\bar X+\frac{1+2\Xi}{2} =\mathcal H-b_0\bar R.
 \label{eq:charge_cancellation}
\end{align}
Here we used $r_{,\rho}/r=1/2$, $\nu_{,\rho}=\bar A$, $(f')_{,\rho}/f'=\delta\bar X$, and $\bar K_{,\rho}/\bar K=-(1+2\Xi)$. Consequently,
\begin{align}
 \mathscr Q&=\mathcal P S =r^2e^\nu\left(b_0f'A+b_1\widehat{f'}\right), \label{eq:radial_charge}\\
 \mathscr Q_{,\rho} &=\mathcal P\left[S_{,\rho}+(\mathcal H-b_0\bar R)S\right]=0. \label{eq:purple_Q_cancellation}
\end{align}
The second expression for $\mathscr Q$ uses the outward radial orientation in the metric dictionary, for which $\phi>0$. This proof does not divide by $S$ and therefore includes the invariant zero-charge sector $S=0$. For $b_0\ne0$, that sector also satisfies $e^\nu|f'|^{b_1/b_0}=\mathrm{constant}$. The charge depends on the normalization of the Killing time. Its value on the full Clifton--Barrow family is evaluated in Sec.~\ref{subsec:clifton_barrow_metric_current}. Related first integrals have been obtained by Noether methods \cite{CapozzielloStabileTroisi2007Noether}.

\subsection{Schwarzschild as a Ricci-flat invariant orbit}
\label{subsec:schwarzschild_metric_orbit_current}

For positive $\delta$, $f(0)=f'(0)=0$. Ricci-flat metrics satisfy the undivided vacuum equations, but the curvature-fluid expressions are undefined there. We select the GR-like Ricci-flat extension by imposing vanishing geometrical source and
\begin{equation}
 \Gamma_{\rm Schw}:\qquad \bar R=\bar Z=0,\qquad \bar K=\bar A+\frac14.
 \label{eq:Schwarzschild_orbit_current}
\end{equation}
The last condition distinguishes this orbit from the wider scalar-flat boundary. Along it, Eq.~\eqref{eq:A_red_R0} becomes
\begin{equation}
 \bar A_{,\rho}=-\bar A\left(2\bar A+\frac12\right).
 \label{eq:Abar_Schwarzschild_orbit}
\end{equation}
Integration and the metric dictionary yield
\begin{align}
 \bar A(r)&=\frac{M}{2(r-2M)}, \label{eq:Abar_Schwarzschild_metric}\\
 \bar K(r)&=\frac{1}{4(1-2M/r)}, \label{eq:Kbar_Schwarzschild_metric}\\
 ds^2&=-C_t\left(1-\frac{2M}{r}\right)dt^2 +\frac{dr^2}{1-2M/r}+r^2d\Omega^2. \label{eq:Schwarzschild_metric}
\end{align}
For $M>0$, this static chart covers $r>2M$; choosing $C_t=1$ fixes the usual asymptotic normalization. The case $M=0$ is Minkowski. Every finite-mass member approaches $(\bar A,\bar R,\bar Z,\bar K)=(0,0,0,1/4)$ as $r\to\infty$. Its representative $P_{\rm S}$ in the polynomial extension has $\bar X=0$, a label for this selected extension rather than the value of a defined quotient on $R=0$.

A variation within the Schwarzschild family gives
\begin{equation}
 \Delta_M\bar A=\Delta_M\bar K =\frac{r\,\Delta M}{2(r-2M)^2} \sim\frac{\Delta M}{2r}.
\end{equation}
This outward decay explains attraction along the family. The formal $\bar R$-normal variational equation of the polynomial extension, evaluated along its $\bar X=0$ representative, is
\begin{equation}
 \Delta\bar R_{,\rho}=(1-2\bar A)\Delta\bar R.
 \label{eq:DeltaR_Schwarzschild_current}
\end{equation}
Using $d\rho=2\,dr/r$, its solution is
\begin{equation}
 \Delta\bar R=D\frac{r^3}{r-2M}\sim Dr^2.
\end{equation}
For $M>0$, its magnitude decreases between $2M$ and $3M$ and grows beyond $3M$. Since the background has $\phi^2=4(r-2M)/r^3$, the corresponding formal dimensional mode is $\Delta R=\phi^2\Delta\bar R=4D$. It must vanish for an orbit to approach the selected Schwarzschild endpoint. This is a variational statement about the polynomial extension, not a regular linearization of the metric field equations at $f'(0)=0$. The remaining stable $\bar X$ direction is likewise chart dependent and does not define an independent perturbation of an exact Ricci-flat metric.

\subsection{The Clifton--Barrow family and its fixed point}
\label{subsec:clifton_barrow_metric_current}

Let $\bar A_{\rm C}$, $\bar R_{\rm C}$, and $\bar K_{\rm C}$ denote the values at $P_{\rm C}$. The full Clifton--Barrow family \cite{CliftonBarrow2005,Clifton:2006ug} can be written as
\begin{equation}
 ds^2=-C_t\left(\frac r{r_0}\right)^{p_\delta}h(r)\,dt^2 +\frac{4\bar K_{\rm C}}{h(r)}\,dr^2+r^2d\Omega^2,
 \label{eq:purple_CB_full}
\end{equation}
where
\begin{equation}
\label{eq:purple_CB_h}
h(r)=1+C_{\rm CB}\left(\frac r{r_0}\right)^{-n_\delta}, \qquad p_\delta=\frac{2\delta(1+2\delta)}{1-\delta}.
\end{equation}
Here $n_\delta=N_\delta/(1-\delta)$, with $N_\delta$ defined after Eq.~\eqref{eq:PC_eigenvalues}. The $C_{\rm CB}$ is dimensionless, $r_0>0$ is held fixed, and $C_t>0$ fixes the time normalization. We consider static patches with $h>0$ and $\bar K_{\rm C}>0$. For $0<\delta<\delta_{\rm K}$, the latter condition holds and $|g_{tt}|/(r/r_0)^{p_\delta}\to C_t$ at large radius. Choosing $C_t=1$ fixes this leading power-law coefficient.

The curvature and normalized variables are
\begin{align}
 R(r)&=\frac{\kappa_\delta}{r^2},\qquad \kappa_\delta=\frac{\bar R_{\rm C}}{\bar K_{\rm C}} =-\frac{6\delta(1+\delta)}{1-2\delta-2\delta^2}, \label{eq:purple_CB_curvature}\\
 \bar A&=\bar A_{\rm C}+\frac{rh_{,r}}{4h},\qquad \bar K=\frac{\bar K_{\rm C}}h,\qquad \bar X=-1. \label{eq:purple_CB_normalized}
\end{align}
Thus, $C_{\rm CB}=0$ gives precisely the scale-invariant equilibrium $P_{\rm C}$, since $p_\delta=4\bar A_{\rm C}$. The identity \eqref{eq:Rbar_metric_AK_relation} then reproduces $\bar R_{\rm C}$, and $\bar X=-1$ agrees with Eq.~\eqref{eq:Xbar_minus_one_selfsimilar}. For this member, $r=0$ is a curvature singularity and there is no positive-radius Killing horizon. The horizon structure of members with $C_{\rm CB}\ne0$ is not fixed by the equilibrium alone.

The charge \eqref{eq:radial_charge} measures the integration constant once the normalization is fixed. In fact,
\begin{align}
 f'(r)&=F_0\left(\frac r{r_0}\right)^{-2\delta},\qquad F_0=(1+\delta)\left|\frac{\kappa_\delta}{R_0r_0^2}\right|^\delta, \label{eq:purple_CB_fprime}\\
 S&=\frac{b_0rh_{,r}}{4h}.
\end{align}
Using $1+p_\delta/2-2\delta=n_\delta$ gives
\begin{equation}
 \mathscr Q=-\frac{b_0n_\delta}{4\sqrt{\bar K_{\rm C}}} r_0\sqrt{C_t}\,F_0C_{\rm CB}.
 \label{eq:purple_CB_charge}
\end{equation}
For the alternative convention $h=1+C/r^{n_\delta}$, substitute $C_{\rm CB}=C/r_0^{n_\delta}$ in this expression. The sign follows from the plus sign in \eqref{eq:purple_CB_h}. The equilibrium has zero charge; with $C_t=1$ and $C_{\rm CB}\to-2M/r_0$, the GR limit gives $\mathscr Q\to M/3$.

The radial spectrum also has a direct metric interpretation. Equation~\eqref{eq:PC_eigenvalues} gives
\begin{equation}
 e^{\lambda_1\rho}=\left(\frac r{r_0}\right)^{-n_\delta},
\end{equation}
the relative falloff of the $C_{\rm CB}$ term. When $\Delta_{\rm C}<0$, the other two eigenvalues give the real modes
\begin{equation}
 \left(\frac r{r_0}\right)^{-n_\delta/2} \left\{\cos\!\left[\omega_{\rm C}\ln\frac r{r_0}\right], \sin\!\left[\omega_{\rm C}\ln\frac r{r_0}\right]\right\},
\end{equation}
where $\omega_{\rm C}=\sqrt{-\Delta_{\rm C}}/[2(1-\delta)]$. These reproduce the radial modes of the direct metric analysis \cite{CliftonBarrow2005}. The sink and focus-to-node classification over $0<\delta<\delta_{\rm K}$ was established in Sec.~\ref{out_lin_stab}; it concerns static radial asymptotics.

Finally, the GR limit is transparent in variables that remain defined on scalar-flat metrics:
\begin{equation}
 (\bar A_{\rm C},\bar R_{\rm C},\bar Z_{\rm C},\bar K_{\rm C}) \longrightarrow(0,0,0,1/4),\qquad \bar Z_{\rm C}=-\bar R_{\rm C}.
 \label{eq:PC_to_PS_regular_limit_current}
\end{equation}
The fixed-point metric therefore tends to Minkowski after time normalization. Keeping $C_{\rm CB}$ before taking $\delta\to0$ instead recovers the full Schwarzschild family. The fixed value $\bar X_{\rm C}=-1$ causes no contradiction, since $\bar X$ is undefined on the limiting exact scalar-flat metric.

\subsection{Degenerate fixed sets and selected scalar-flat orbits}
\label{subsec:other_fixed_points_metric_current}

The remaining finite critical sets satisfy
\begin{equation}
 \left.\bar K\right|_{\mathcal L_0(\sigma)}=0, \qquad \bar K(P_{K0})=0.
 \label{eq:Kbar_zero_finite_sets_current}
\end{equation}
By \eqref{eq:Kbar_metric_dictionary}, these sets themselves do not give nondegenerate metrics in the static areal-radius chart. They instead organize limiting orbits. At $\delta=1/4$, $P_{K0}$ joins $\mathcal L_0$; at $\delta=\delta_{\rm K}$, it coalesces with $P_{\rm C}$. Its changes of radial stability do not make it a regular spherical equilibrium metric.

Positive-$\bar K$ orbits approaching $\mathcal L_0$ can nevertheless be reconstructed. On a nonstationary line $\bar X=c\bar A$ of the polynomially extended plane, set $q=\delta c$ and use $\bar K(\bar A)$ from \eqref{eq:R0_line_flow}. The metric dictionary gives
\begin{equation}
 \frac{d\ln r}{d\bar A}=-\frac{1}{4\bar A\bar K}, \qquad \frac{d\nu}{d\bar A}=-\frac{1}{2\bar K}.
 \label{eq:purple_scalarflat_quadratures}
\end{equation}
These quadratures apply on intervals with $\bar A\ne0$ and $\bar K>0$. The line $c=0$ yields Schwarzschild. The separate axis $\bar A=0$ gives $\nu=\nu_0$ and $g^{rr}=1-C/r$.

For $q>0$, define
\begin{equation}
 \Delta=\sqrt{1+q+q^2},\qquad \bar A_\pm=\frac{-(1+q)\pm\Delta}{2q}.
 \label{eq:scalarflat_roots}
\end{equation}
Then $\bar K=q(\bar A-\bar A_-)(\bar A-\bar A_+)$ and $\bar A_-<-1<\bar A_+<0$. The positive-$\bar K$ component $\bar A<\bar A_-$ is disconnected from $P_{\rm S}$ within the finite normalized chart. Appendix~\ref{app:purple_scalarflat} gives the integrated metric \eqref{eq:R0_disconnected_reconstruction} and verifies both its scalar flatness and its endpoint geometry. As $\bar A\to-\infty$, the metric extends smoothly through a minimal sphere with positive lapse. The extension reaches an asymptotically flat end on the branch $\bar A>0$; for $q>0$ it is not Schwarzschild. The throat realizes a physical direction on $\mathscr C_{AX}$.

At the other end, $\bar A\to\bar A_-^-$ and $r\to\infty$, while $g^{rr}\sim r^\gamma$ with $\gamma=-4\bar A_-\Delta>4$. The angular orthonormal curvature component $(1-g^{rr})/r^2$ diverges. Radial null geodesics reach this end in finite affine parameter, and no intervening finite-radius
Killing horizon shields it. Thus the smooth extension has one asymptotically flat end and one curvature-singular end.

This family is identifiable with the special Buchdahl-inspired metric \cite{Nguyen2023Buchdahl}. The parameter correspondence is $\widetilde k=-q/(q+2)$, so $q>0$ maps to $-1<\widetilde k<0$; Appendix~\ref{app:purple_scalarflat} derives the identification. The two-ended construction discussed in Ref.~\cite{NguyenAzregAinou2023Wormholes} uses an additional gluing prescription, whereas the smooth continuation considered here retains the singular end. More generally, scalar-flat non-Schwarzschild geometries with throats were studied in Ref.~\cite{DadhichEtAl2002ScalarFlat}. Our result identifies this selected family within the covariant boundary flow and connects its throat and singular end to the compactified and finite critical sets.

These reconstructed metrics satisfy the undivided vacuum equations for positive $\delta$, since $R\equiv0$ makes $f=f'=0$ and the composite field $f'(R)$ constant. They do not exhaust all scalar-flat vacua: $\bar X$ is undefined there, so the first integral \eqref{eq:R0_first_integral} cannot constrain every exact scalar-flat metric.

\subsection{Metric interpretation of the asymptotic critical sets}\label{subsec:asymptotic_metric_reconstruction_current}

\begin{table*}[ht]
\caption{Metric information supplied by the finite and resolved asymptotic critical sets. Projective data alone do not fix a metric.}
\label{tab:metric_summary}
\small
\setlength{\tabcolsep}{4pt}
\begin{tabular}{||c|c|c||}
\hline\hline
Critical set & Metric data & Interpretation \\
\hline
$P_{\rm S}$, $\Gamma_{\rm Schw}$ & $\bar R=\bar Z=0$; selected orbit has $\bar K=\bar A+1/4$ & scalar-flat approaches $P_{\rm S}$, including Schwarzschild. \\[3pt]
$P_{\rm C}$ & $\bar R_{\rm C}\ne0$, $\bar X_{\rm C}=-1$ & Scale-invariant Clifton--Barrow member. \\[3pt]
$\mathcal L_0$, $P_{K0}$ & $\bar K=0$ & Degenerate equilibrium metrics. \\[3pt]
$\mathscr C_{AX}$ & $\bar K\sim\delta u_Au_X/z^2$ & No unique metric; (regular-throat example). \\[3pt]
$Q_A^\epsilon,Q_X^\epsilon,Q_{RX}^\epsilon$ & Secondary charts give distinct lifts & Partial local classification. Subleading data needed. \\[3pt]
$\mathscr Q_R^\epsilon$ & $\bar K\sim D_\epsilon/w^2$, $Rr^2\to\epsilon/D_\epsilon$ for $D_\epsilon>0$ & Partial metric data; $D_\epsilon=0$ requires subleading analysis. \\
\hline\hline
\end{tabular}
\end{table*}

Phase space infinity means that at least one of the normalized variables \((\bar A,\bar R,\bar X)\) diverges. It does not necessarily correspond to \(r\rightarrow\infty\); it may instead describe a finite-radius singularity, a horizon-like boundary, or a degeneration of the normalization by \(\phi\). Consequently, a projective direction does not by itself determine a unique spacetime metric.

For the normally hyperbolic circle \(\mathscr C_{AX}\), the results of Sec.~\ref{subsec:compactification_infinity_reduced} give
\[
\bar K\sim\frac{\delta u_Au_X}{z^2}, \qquad g_{rr}=4\bar K, \qquad Rr^2=\frac{\bar R}{\bar K}\longrightarrow0.
\]
For $\delta>0$, compatibility with $\bar K>0$ requires $u_Au_X>0$, and then $g_{rr}\to+\infty$. This is a failure of the areal-radius chart at a possible throat, horizon, or singular boundary; the orbit must be reconstructed to distinguish them. The scalar-flat family above supplies an explicit smooth-throat realization. The physical part of $\mathscr C_{AX}$ repels in the compactifying normal direction.

The secondary charts distinguish different lifts of $Q_A^\epsilon$ and $Q_X^\epsilon$, as well as the mixed ray $Q_{RX}^\epsilon$. Their phase-space spectra do not by themselves determine unique metric representatives. The metric reconstruction still requires the orbit-dependent subleading relations, particularly when the leading term of the Gaussian-curvature constraint vanishes. Divergence regimes not covered by these charts require additional local blow-ups.

For the weighted sheet \(\mathscr Q_R^\epsilon\), define
\[
D_\epsilon(U,V) \equiv \delta UV+\epsilon\frac{\delta}{2(1+\delta)}.
\]
When \(D_\epsilon\neq0\), the leading metric information is
\[
\bar K =\frac{D_\epsilon}{w^2}+O(w^{-1}), \qquad g_{rr}\sim\frac{4D_\epsilon}{w^2}, \qquad Rr^2\longrightarrow\frac{\epsilon}{D_\epsilon}.
\]
Thus \(D_\epsilon>0\) is necessary for compatibility with a Lorentzian areal-radius chart. These relations fix only a curvature ratio and the leading radial metric coefficient; the functions \(g_{tt}(r)\) and \(R(r)\) still depend on the orbit-dependent approach of \(w(r)\), \(U(r)\), and \(V(r)\). If \(D_\epsilon=0\), the leading term vanishes and a further blow-up is required.

For \(0<\delta<1\), only the positive-\(\bar R\) sheet contains a normally attracting subset compatible with \(\bar K>0\). The negative-\(\bar R\) sheet contains no such subset. This is a large-normalized-curvature regime, not necessarily a large physical-curvature regime, and it is dynamically separated from the negative-\(\bar R\) Clifton--Barrow equilibrium by the invariant boundary \(\bar R=0\).

The metric interpretation of the finite and asymptotic critical sets is summarized in Table~\ref{tab:metric_summary}.

\section{Concluding remarks}\label{sec:conclusion}

We have formulated the static spherical vacuum equations of signed power-law metric $f(R)$ gravity as a covariant autonomous system and connected its critical sets to metric reconstructions. The regular finite attractor for $0<\delta<\delta_{\rm K}$ is the scale-invariant Clifton--Barrow equilibrium in the negative-curvature sector. Its spectrum reproduces the known static radial modes, while the radial first integral is proportional to the integration constant of the full Clifton--Barrow family after the lapse normalization is fixed. The selected Schwarzschild extension instead approaches a Minkowski endpoint that is a saddle of the polynomially extended system.

The compactification resolves several additional balances. The physical part of $\mathscr C_{AX}$ is normally saddle-like and repels in the compactifying direction. Full secondary charts distinguish different lifts of the degenerate rays and leave specific zero modes unresolved. The weighted curvature charts yield critical sheets whose neutral directions retain orbit-dependent metric data. For $0<\delta<1$, only the positive-$\bar R$ sheet contains a normally attracting subset compatible with the spherical sector. The invariant boundary $\bar R=0$ separates that subset from the Clifton--Barrow equilibrium within the regular chart.

Reconstruction of selected scalar-flat boundary orbits identifies the $q>0$ family with the special Buchdahl-inspired metric \cite{Nguyen2023Buchdahl,NguyenAzregAinou2023Wormholes}. Its smooth continuation connects a regular throat to one asymptotically flat end and one null-incomplete curvature-singular end. The finite critical curve and the circle at infinity thus have different geometric realizations along a single extended metric. This identification complements the earlier coordinate-based and Einstein-frame analyses \cite{MignemiWiltshire1989,CliftonBarrow2005} by locating their geometries within the present covariant phase portrait.

The scope of these conclusions is limited in three respects. First, radial stability does not establish stability under time-dependent perturbations; in particular, $f''<0$ on the negative-curvature branch for $0<\delta<1$. Second, $f'(0)=0$ makes the exact scalar-flat sector degenerate, so the selected boundary reconstruction cannot classify all such vacua. This differs from the nondegenerate case $f(0)=0$, $f'(0)\ne0$, in which the scalar-flat spherical vacuum equations select Schwarzschild \cite{Nzioki:2009av}. Third, exceptional parameters and the unresolved directions at infinity require separate analyses. The GR limit likewise requires care: the fixed-point metric tends to Minkowski, while the full Schwarzschild family is recovered by retaining the integration constant before taking the limit.

Further work can address the remaining center directions, time-dependent perturbations, and matching to material interiors with the full $f(R)$ junction conditions \cite{RosaCarloni2024}. These extensions would determine which static radial asymptotics can serve as exteriors of physically admissible compact objects.

\section*{Acknowledgments}
EB is partially supported by \textit{Conselho Nacional de Desenvolvimento Científico e Tecnológico} (grant N.\ 306305/2026-7) and FAPEMIG (APQ-05207-23). PKSD is supported by a grant from First Rand Bank (SA).

\appendix

\section{Verification and identification of the scalar-flat reconstruction}
\label{app:purple_scalarflat}

By using $\bar K(\bar A)$ from \eqref{eq:R0_line_flow} and the roots \eqref{eq:scalarflat_roots}, Eqs.~\eqref{eq:purple_scalarflat_quadratures} and \eqref{eq:areal_radius_metric} give
\begin{equation}
 \bar A_{,r}=-\frac{4\bar A\bar K}{r}, \qquad \nu_{,r}=\frac{2\bar A}{r}, \qquad e^{2\lambda}=4\bar K.
 \label{eq:purple_app_flow}
\end{equation}
The integrated solution is
\begin{align}
 r(\bar A)&=r_{\rm th} \left|1-\frac{\bar A_-}{\bar A}\right|^{1/(4\Delta\bar A_-)} \left|1-\frac{\bar A_+}{\bar A}\right|^{-1/(4\Delta\bar A_+)}, \nonumber\\
 e^{\nu(\bar A)}&=e^{\nu_{\rm th}} \left|\frac{\bar A-\bar A_-}{\bar A-\bar A_+}\right|^{1/(2\Delta)},
 \label{eq:R0_disconnected_reconstruction}
\end{align}
with $r_{\rm th}>0$. We first verify the local curvature and the smooth extension, then identify the family with the Buchdahl-inspired metric.

\subsection{Direct curvature calculation}

Set $B=g^{rr}=1/(4\bar K)$. For the static spherical metric,
\begin{align}
 R=\frac{2}{r^2}-2B\bigg[&\nu_{,rr}+\nu_{,r}^2 -\nu_{,r}\lambda_{,r}\nonumber\\
 &+\frac{2}{r}(\nu_{,r}-\lambda_{,r})+\frac{1}{r^2}\bigg].
 \label{eq:purple_app_R}
\end{align}
Substitution of \eqref{eq:purple_app_flow} gives
\begin{equation}
 Rr^2=2- \frac{1+2\bar A+4\bar A^2-8\bar A\bar K +4\bar A(\bar A+1)\bar K_{,\bar A}}{2\bar K}=0.
 \label{eq:purple_app_R_zero}
\end{equation}
The numerator is identically $4\bar K$ for the quadratic $\bar K(\bar A)$ in \eqref{eq:R0_line_flow}. The metric is generally not Ricci flat, since
\begin{equation}
 R_{\hat t\hat t}=-\frac{q\bar A^2}{\bar K r^2}\ne0 \qquad (q>0,\ \bar A\ne0).
 \label{eq:purple_app_Rtt}
\end{equation}
For positive $\delta$, $R\equiv0$ makes the composite field $f'(R)=0$ constant. Thus the undivided vacuum equations \eqref{eq:fR_field_eqs} vanish, without evaluating the singular curvature-fluid quotients.

\subsection{Regular throat and extension}

The inner end of $\bar A<\bar A_-$ has $\bar A\to-\infty$. Introduce
\begin{equation}
 x=\frac1{\bar A},\qquad \mathcal D(x)=q+(1+q)x+\frac{x^2}{4},\qquad \bar K=\frac{\mathcal D(x)}{x^2}.
 \label{eq:purple_app_s}
\end{equation}
Then
\begin{equation}
 \frac{d\ln r}{dx}=\frac{x}{4\mathcal D(x)}, \qquad \frac{d\nu}{dx}=\frac1{2\mathcal D(x)},
 \label{eq:purple_app_s_flow}
\end{equation}
and the metric becomes
\begin{equation}
 ds^2=-e^{2\nu(x)}dt^2 +\frac{r(x)^2}{4\mathcal D(x)}\,dx^2+r(x)^2d\Omega^2.
 \label{eq:purple_app_regular_metric}
\end{equation}
Since $\mathcal D(0)=q>0$, the equations and metric coefficients are analytic near $x=0$, with $r(0)=r_{\rm th}$ and $e^{\nu(0)}=e^{\nu_{\rm th}}>0$. Choose proper distance by $d\ell/dx=r/(2\sqrt{\mathcal D})$, $\ell(0)=0$. Direct differentiation gives
\begin{equation}
 r_{,\ell}(0)=0,\qquad r_{,\ell\ell}(0)=\frac1{r_{\rm th}},\qquad \nu_{,\ell}(0)=\frac1{r_{\rm th}\sqrt q}.
 \label{eq:purple_app_throat_data}
\end{equation}
Hence
\begin{equation}
 r(\ell)=r_{\rm th}+\frac{\ell^2}{2r_{\rm th}}+O(\ell^3).
 \label{eq:purple_app_throat_series}
\end{equation}
The sphere is a regular minimum of area with positive lapse. The $x>0$ extension has $\bar A>0$; as $\bar A\to0^+$, $r\to\infty$, $4\bar K\to1$, and $\bar A=O(r^{-1})$. The lapse tends to a positive constant, removable by a rescaling of $t$, so this end is asymptotically flat. Equation~\eqref{eq:purple_app_Rtt} shows that it is not Schwarzschild for $q>0$.

\subsection{Curvature-singular end and null incompleteness}

At the opposite end, put $\varepsilon=\bar A_--\bar A>0$. As $\varepsilon\to0^+$,
\begin{align}
 \bar K&=\Delta\varepsilon+O(\varepsilon^2),\nonumber\\
 r&\sim r_*\varepsilon^{1/(4\Delta\bar A_-)},\qquad e^\nu\sim N_*\varepsilon^{1/(2\Delta)},
 \label{eq:purple_app_outer_eps}
\end{align}
where $r_*,N_*>0$. Equivalently,
\begin{equation}
 B\sim B_*r^\gamma,\quad e^\nu\sim\widetilde N_*r^{2\bar A_-}, \quad \gamma=-4\bar A_-\Delta>4.
 \label{eq:purple_app_outer_r}
\end{equation}
The angular orthonormal component is $R_{\hat\theta\hat\varphi\hat\theta\hat\varphi}=(1-B)/r^2$. For this static diagonal metric, the Kretschmann scalar is a sum of nonnegative squared orthonormal components, so
\begin{equation}
 R_{abcd}R^{abcd}\geq4\left(\frac{1-B}{r^2}\right)^2 \longrightarrow+\infty.
 \label{eq:purple_app_Kretschmann}
\end{equation}
For a radial null geodesic with conserved energy $E=e^{2\nu}dt/d\lambda_{\rm aff}>0$,
\begin{align}
 \left|\frac{d\lambda_{\rm aff}}{dr}\right| &=\frac{e^\nu}{E\sqrt B} \sim r^{2\bar A_-(1+\Delta)}, \label{eq:purple_app_affine}\\
 \left|\frac{dt}{dr}\right| &=\frac{e^{-\nu}}{\sqrt B} \sim r^{-(q+\Delta)}. \label{eq:purple_app_time}
\end{align}
Both integrals to $r=\infty$ converge: the first exponent is below $-1$, and $q+\Delta>1$. The singular end is null incomplete. The lapse is positive at every interior point, including the throat, so no intervening finite-radius Killing horizon shields it.

\subsection{Identification with the Buchdahl-inspired metric}

An explicit comparison fixes the relation to Refs.~\cite{Nguyen2023Buchdahl,NguyenAzregAinou2023Wormholes}. On the static branch $0<y<1$, their metric can be written as
\begin{align}
 ds^2={}&-y^{a_{\rm B}}dt^2  +L_{\rm B}^2\frac{y^{b_{\rm B}}}{(1-y)^4}\,dy^2+L_{\rm B}^2\frac{y^{b_{\rm B}+1}}{(1-y)^2}\,d\Omega^2, \label{eq:Buchdahl_comparison_metric}\\
 a_{\rm B}&=\frac{1+\widetilde k}{\zeta_{\rm B}},\qquad b_{\rm B}=\frac{\widetilde k-1}{\zeta_{\rm B}},\qquad \zeta_{\rm B}=\sqrt{1+3\widetilde k^2}. \label{eq:Buchdahl_parameters}
\end{align}
Here $L_{\rm B}>0$ sets the radial scale and a constant time rescaling is understood. The areal radius is $r=L_{\rm B}y^{(b_{\rm B}+1)/2}/(1-y)$. Writing $J_{\rm B}=(b_{\rm B}+1)+(1-b_{\rm B})y$, the dictionary gives
\begin{equation}
 \bar A=\frac{a_{\rm B}(1-y)}{2J_{\rm B}},\qquad \bar K=\frac{y}{J_{\rm B}^2}.
\end{equation}
Eliminating $y$ yields
\begin{equation}
 \bar K=\frac14-\frac{b_{\rm B}}{a_{\rm B}}\bar A +\frac{b_{\rm B}^2-1}{a_{\rm B}^2}\bar A^2.
 \label{eq:Buchdahl_normalized_constraint}
\end{equation}
The parameter choice
\begin{equation}
 \widetilde k=-\frac{q}{q+2},\qquad a_{\rm B}=\frac1\Delta,\qquad b_{\rm B}=-\frac{1+q}{\Delta}
 \label{eq:Buchdahl_parameter_map}
\end{equation}
makes \eqref{eq:Buchdahl_normalized_constraint} exactly $\bar K=1/4+(1+q)\bar A+q\bar A^2$. Direct differentiation also gives $r\bar A_{,r}=-4\bar A\bar K$, so the reconstruction equations coincide up to the radial and time normalization constants. Thus $q>0$ corresponds to the known range $-1<\widetilde k<0$. The branch $0<y<1$ crosses its regular minimum of areal radius and retains the singular end at $y\to0$. The further gluing used to construct two asymptotically flat ends in Ref.~\cite{NguyenAzregAinou2023Wormholes} is a separate global prescription, not part of this continuation.

\bibliography{bibliography}

\end{document}